\documentclass[aps,
twocolumn,superscriptaddress,showpacs]{revtex4}
\usepackage{epsf}
\usepackage{graphicx}
\usepackage{amsmath}
\usepackage{psfrag}
\begin{document}
\title{Polymers in crowded environment under stretching force: globule-coil transitions}
\author{Viktoria Blavatska}
\email[]{E-mail: viktoria@icmp.lviv.ua; blavatska@itp.uni-leipzig.de}
\affiliation{Institut f\"ur Theoretische Physik and Centre for Theoretical Sciences (NTZ),\\ Universit\"at Leipzig, Postfach 100\,920,
D-04009 Leipzig, Germany}
\affiliation{Institute for Condensed
Matter Physics of the National Academy of Sciences of Ukraine,\\
79011 Lviv, Ukraine}
\author{Wolfhard Janke}
\email[]{E-mail: Wolfhard.Janke@itp.uni-leipzig.de}
\affiliation{Institut f\"ur Theoretische Physik and Centre for Theoretical Sciences (NTZ),\\ Universit\"at Leipzig, Postfach 100\,920,
D-04009 Leipzig, Germany}
 
\begin{abstract}
We study flexible polymer macromolecules in a crowded (porous) environment, modelling them as self-attracting self-avoiding 
walks on site-diluted percolative lattices 
in space dimensions $d=2$, $3$.
The influence of stretching force on the polymer folding and properties of globule-coil transitions are analyzed. 
Applying the pruned-enriched Rosenbluth chain-growth method (PERM),
we estimate the transition temperature $T_{\Theta}$ between collapsed and extended polymer configurations
and construct the phase diagrams of the globule-coil coexistence when varying temperature and stretching force. 
The transition to a completely stretched state, caused by applying force, is discussed as well.

\end{abstract}
\maketitle

\section{Introduction}
Long flexible polymer macromolecules in a good solvent possess configurational statistics, which is perfectly captured by  
the model of self-avoiding random walks (SAW) on a regular lattice  \cite{polymerbook}. This corresponds to the  {\it regime of polymer coils} which holds provided that
the temperature $T$ is above the so-called $\Theta$-temperature. In this regime, the mean end-to-end distance of an $N$-step chain 
scales as $R_N\sim N^{\nu_{{\rm SAW}}}$, where $\nu_{{\rm SAW}}>1/2$ is an universal exponent, which depends on space dimension $d$ only. With lowering the temperature, the effect of monomer-monomer attraction grows and the polymer radius shrinks. At $T=T_{\Theta}$ the effective repulsion due to the volume exclusion constraint is exactly balanced  
by attractive interactions. At this particular temperature a crossover occurs from high temperature SAW behavior to  
$\Theta$-statistics with exponent $\nu_{\Theta}(d=2)=4/7$ \cite{Duplantier82}, $\nu_{\Theta}(d\geq3)=1/2$.  
Below the $\Theta$-temperature, the entropic effects, which make the polymer chain swell, are overcome by interaction energy and   
a collapse to the {\it globule regime} (with collapsed size exponent $\nu_{c}=1/d$) occurs.  
It is generally accepted \cite{polymerbook} that the coil-globule 
transition is of second order, in the sense that the density of an infinite globule is zero at $T=T_{\Theta}$ and increases continuously when $T$ is 
lowered further; more presicely, it is  a tricritical point with the upper critical dimension $d_c=3$.

The coil-globule transition is one of the fundamental problems in polymer physics \cite{polymerbook,Grosberg94, Vanderzande98}, being deeply 
connected with problems like protein folding, DNA condensation and chromatin organization. 
The properties of polymers in the vicinity of the $\Theta$-point can be successfully studied on the basis of self-attractive self-avoiding walks (SASAW), where a 
nearest-neighbor interaction is included.
Estimates of the transition temperature $T_{\Theta}$
for flexible polymers  have been so far the subject of 
numerous studies \cite{Privman86,Lam90,Foster92,Barkema98,Szleifer92,Grassberger95,Grassberger97,Vogel07}.   
Note that the $\Theta$-temperature is a non-universal quantity, depending on the lattice type; in particular it appears to be proportional to the connectivity constant $\mu$ of a SAW on a given type of lattice (recent estimates of $\mu$ and $\Theta$-temperatures  on simple square and cubic lattices are given in Table 1).


\begin{table}[b!]
\small{
\caption{ Values of connectivity constants and $\Theta$-temperatures for SASAWs on pure regular lattices and site-diluted 
 percolative lattices for different space dimensions $d$.}
\label{dim}
\begin{center}
\begin{tabular}{lcccc}
\hline
\hline  
 $d$ & $\mu$ & $T_{\Theta}$ & $\mu_{p_c}$ & $T_{\Theta}^{p_c}$ (our study) \\
\hline 
2 & 2.6385(1) \cite{Guttmann91} & 1.499(2) \cite{Barkema98} & 1.565(2) \cite{Ordemann00} &0.92(2) \\
3 & 4.68404(9) \cite{MacDonald00}& 3.717(3) \cite{Grassberger97}& 1.462(2) \cite{Ordemann00} &0.71(2)\\
 \hline 
\hline
\end{tabular}
\end{center}
}
\end{table}

 The recent progress in experimental techniques employing optical tweezers \cite{Bockelmann02}, atomic force microscopy \cite{Rief99} and soft
 microneedles \cite {Kishino88} makes it possible to monitor the behavior of various polymers under tension and stress. In particular, applying a force on 
 an isolated protein, the mechanism of force-driven phase transitions was studied, such as unfolding of giant titine protein \cite{Rief97}, 
 the stretching and unzipping of collapsed DNA molecules \cite{Baumann00}.
 Of special interest in biophysics is the stretching of a collapsed polymer, i.e.,  of a polymer in a poor solvent below the $\Theta$-temperature. Unfolding proteins 
in this way could give important information on their spontaneous folding pathways. An intriguing question is whether intermediate stages observed in different protein denaturating processes under applying force would always correspond to folding pathways in an enforced environment \cite{Lu98,Bryant00,Lemak03}.
Note that varying temperature alone, a polymer cannot acquire the conformation of the completely stretched state with size exponent  $\nu=1$. Hence force 
not only influences the elastic, mechanical and structural properties of polymers,
 but also introduces a new stretched state ({\it regime of string}) which is otherwise not accessible. 
 The properties of force-induced globule-coil and coil-string transitions in $d=2$, $3$ have been studied intensively both analytically \cite{Halperin91,Goritz96,Grassberger02,Marenduzzo02,Ciepak04} and applying numerical simulations \cite{Grassberger02,Kumar}. 
The existence of many intermediate states was found when unfolding under an applied force, which are not the same as the thermodynamically 
stable intermediate states. The phase diagrams of the globule-coil and coil-string coexistence when varying temperature and stretching force were constructed.

New challenges have been raised recently in studying  protein folding in the natural cellular environment \cite{Goodesel91}.
Real biological cells can be described as a very crowded environment built of the biochemical species, 
which occupy a large fraction of the total volume; the ``volume exclusion" arises due to the steric repulsion between molecules \cite{Minton01}. 
In the language of lattice models, the crowded environment with structural obstacles can be considered as a disordered lattice, where 
some amount of randomly chosen sites contains defects.
Of particular interest is the case, when the concentration $p$ of lattice sites allowed for the SAWs
 equals the critical concentration $p_{c}$ ($p_c(d{=}2)=0.592746$ {\cite{Ziff94}}, $p_c(d{=}3)=0.31160$ {\cite{Grassberger92}})
 and the lattice becomes percolative. 
An incipient percolation cluster appears in the system, on which infinitely long  SAWs can reside.  
In this regime,  the scaling law for the end-to-end distance holds with a new exponent $\nu_{p_c}>\nu_{{\rm SAW}}$ 
\cite{Woo91,Grassberger93,Lee96,Rintoul94,Ordemann00,Blavatska04,Janssen07,Blavatska08}. A hint
to the physical understanding of this phenomenon is given by the
fact that the percolation cluster itself is a fractal object with fractal dimension $d_{p_c}^F$ dependent on $d$. 
In this way, scaling law exponents of SAWs  change  with the
dimension $d_{p_c}^F$ of the (fractal) lattice on which the walk
resides. 
Note that studying the scaling of SAWs on a percolative lattice, one is interested rather in the backbone of the percolation cluster,  
which is defined as follows. Assume that each site of the cluster is
a resistor, the neighbour sites are connected by conducting bonds and an external potential drop is applied at two ends of the cluster. The backbone is 
the subset of the cluster consisting of all sites through which the current flows; i.e., it is the structure left when 
all ``dangling ends" are eliminated from the cluster. The SAWs can be trapped in ``dangling ends",  therefore infinitely long chains can 
only exist on the backbone of the cluster, which is characterized by its own fractal dimension $d_{p_c}^B$ ($d_{p_c}^B(d=2)= 1.650\pm0.005$ \cite{Moukarzel98}, $d_{p_c}^B(d=3)=1.86\pm0.01$ \cite{Moukarzel98}).  

Whereas the scaling behavior of SAWs on percolative lattices served as a subject of numerous studies since the early 80th (see, e.g.,
Ref. \cite{Barat95} for a recent review), 
less attention has been paid to peculiarities of the influence of the fractal structure of the underlying lattice on properties of the coil-globule transition.
The upper critical dimension shifts to $d_c=6$ for both SAW and $\Theta$-point statistics on a disordered lattice at the percolation threshold. The scaling
 of polymer size 
at the coil-globule transition point is governed by $\nu_{\Theta}^{p_c}>\nu_{\Theta}$ for $d\leq 6$ (e.g., $\nu_{\Theta}^{p_c}(d=2)=0.74\pm0.02$ \cite{Barat93}, $\nu_{\Theta}^{p_c}(d=3)=0.60\pm0.02$ \cite{Barat93}). 
   It is established that the value of the  $\Theta$-temperature is lowered  due to the presence of disorder
\cite{Barat95,Barat93,Barat91,Roy87,Chang91,Chakrabarti90,Bhattacharya84}, but estimates for $T_{\Theta}^{p_c}$ were found up to now only for the case of 
bond percolation. The existing estimates are: $T_{\Theta}^{p_c}(d=2)=0.62(6)$ \cite{Barat91}, $T_{\Theta}^{p_c}(d=3)=0.43(6)$ \cite {Barat95} (note, that corresponding values of connectivity constant for SAW on bond-diluted percolative lattices read: $\mu_{p_c}\simeq1.29$ and $1.05$ for $d=2$ and 3, respectively \cite {Barat95}). 

The response of a polymer in crowded media to the stretching force modelled by the SASAW model on percolative lattice 
has been considered so far only in $d=2$ for relatively short chains 
by exact enumeration \cite{Kumarpc}. However, much more important is studying this problem in three dimensions, which describes real 
polymer systems.  This  still needs a careful analysis and clarification. Also, a quantitative description of the globule-coil transition 
 under applying force, in particular the estimate of $\Theta$-temperatures in disordered environment under stretching, still remains an open question. 

The aim of the present study is to apply state-of-the-art numerical simulations to analyze the pecularities of SASAWs on site-diluted lattices 
at the percolation threshold (modelling flexible polymers 
in crowded environment) under applied external stretching force in space dimensions $d=2$, $3$. We estimate the shift of the $\Theta$-temperature of 
the globule-coil transition 
under the influence of stretching and analyze the effect of applied force on the phase transitions between collapsed, extended and stretched phases. 

The outline of the rest of the  paper is as follows. In the next section we describe the details of the numerical algorithm used in our study. In section III we present our results of coil-globule transition peculiarities for SAWs on percolative lattices, and in section IV we analyze the 
influence of stretching force on properties of  the transition between collapsed and extended states. We end up by giving conclusions and an outlook in section V.

\section{The method}
We consider site percolation on regular lattices of edge lengths up to $L_{{\rm max}}{=}400,200$ in dimensions $d{=}2,3$, respectively. Each site of 
the lattice was assigned to be occupied with probability $p_c$ and empty otherwise. 
To obtain the backbone of a percolation cluster on a given disordered lattice, we apply an algorithm consisting of the following two steps: first finding the percolation cluster
based on the  site-labeling method of Hoshen and Kopelman \cite{Hoshen76} and then 
 extracting the backbone of this cluster \cite{Porto97} (the algorithm is explained in detail in our previous papers
\cite{Blavatska08,Blavatska09}). We constructed 1000 clusters in each space dimension.

\begin{figure}[b!]
 \begin{center}
\includegraphics[width=4.7cm]{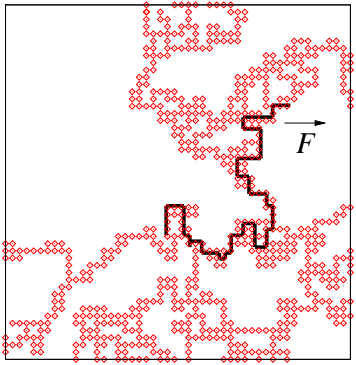}
\end{center}
\caption{\label{sawonpcf} Self-avoding walks on the backbone of a percolation cluster in $d=2$ under stretching force $F$.}
\end{figure}

To study SASAWs on the backbone of percolation clusters, we apply the pruned-enriched Rosenbluth method (PERM) \cite{Grassberger97}, 
taking into account that a SASAW can have its steps only on the sites belonging to the backbone of the percolation cluster.  PERM is based on the original Rosenbluth-Rosenbluth (RR) method 
\cite{Rosenbluth55} and enrichment strategies \cite{Wall59}. The polymer grows step by step, 
i.e., the $n$th monomer is placed at a randomly chosen  empty neighbor site of the last placed $(n-1)$th  monomer ($n\leq N$, where $N$ is the total length of the chain).
The growth is stopped, if the total length of the chain is reached.
In order to obtain correct statistics, any attempt to place a monomer at an already occupied site would result in discarding the entire chain. This leads to an exponential ``attrition" (the number of discarded chains grows exponentially with the chain length). 
The bias due to avoiding this case is corrected in the RR algorithm by means of giving a weight $W_n\sim \prod_{l{=}2}^n m_l$ 
to each sample configuration at the $n$th step, where $m_l$ is the number of free lattice sites to place the $l$th monomer.
This method is particularly useful for studying $\Theta$-polymers, since the Rosenbluth weights of the statistically relevant chains approximately cancel 
against their  Boltzmann probability.  

Population control in PERM suggests pruning configurations with too small weights, and enriching the sample with copies of high-weight 
configurations \cite{Grassberger97}. These copies are made while the chain is growing, and continue to grow independently of each other. Pruning and enrichment are performed by choosing thresholds $W_n^{<}$
and $W_n^{>}$ depending on the estimate of the partition sums of the $n$-monomer chain. If the current weight $W_n$ of an $n$-monomer chain is less than $W_n^{<}$, a random number $r{=}{0,1}$ is chosen; if $r{=}0$, the chain is discarded, otherwise it is kept and its weight is doubled. Thus, low-weight chains are pruned with probability $1/2$. If $W_n$ exceeds  $W_n^{>}$, the configuration is doubled and the weight of each copy is taken as half the original weight. 
For updating the threshold values we apply similar rules as in \cite{Hsu03,Bachmann03}: $W_n^{>}{=}C(Z_n/Z_1)(c_n/c_1)^2$ and $W_n^{<}{=}0.2W_n^{>}$, where $c_n$ denotes the number of created chains having length $n$, and the parameter $C$ controls the pruning-enrichment statistics. 
After a certain number of chains of total length $N$ is produced, the iteration is finished and a new tour starts. 
We adjust the pruning-enrichment control parameter such that on average 10 chains of total length $N$ are generated per each iteration \cite{Bachmann03}, 
and perform $10^6$ iterations.

One end of the chain is subjected to an external force $F$ acting in a chosen direction, say $x$ (see Fig. \ref{sawonpcf}), while the other end (the starting point) is kept fixed. The stretching energy $E_s$ arising due to 
the applied force for an  $n$-step trajectory is given by:
\begin{equation}
E_s=-{\bf F\cdot x}
\label{senergy}
\end{equation}
where ${\bf x}$ is the $x$-component of distance from the starting point $|x_n-x_0|$. 
The Rosenbluth weight factor $W_n$ is thus taken to be:
\begin{equation}
W_n=\prod_{l=2}^n m_l {\rm e}^{\frac{-(E_l-E_{l-1})+{\bf F}|{\bf x}_l-{\bf x}_{l-1}|}{k_B T}},
\label{weight}
\end{equation}
where  $E_l$ denotes the energy of the $l$-step chain ($E_l=z_l\cdot \epsilon$ with  $\epsilon$ being an attractive
energy between two nearest neighbors and $z_l$ the number of nearest neighbors contacts for a given chain) and $k_B$ is the Boltzmann constant.
In what follows, we will assume units in which $\epsilon=-1, k_B=1$.

When a chain of total length $N$  is constructed, a new one starts from the same starting point, until the desired number of chain configurations is obtained.  
The configurational averaging, e.g., for stretching in $x$-direction, is then given by
\begin{eqnarray}
&&\langle x \rangle  
=\frac{\sum_{{\rm conf}} W_N^{{\rm conf}}(x_N^{{\rm conf}}-x_0^{{\rm conf}}) } {\sum_{{\rm conf}} W_N^{{\rm conf}}}, \label{R}
\end{eqnarray}
where $x_0^{{\rm conf}}$ and $x_N^{{\rm conf}}$ denote $x$-coordinates of start- and end-points, respectively,  and $W_N^{{\rm conf}}$ is the weight of an $N$-monomer chain in a given configuration.

Note, that studying SAWs on disordered lattices, we have to perform two types of averaging: the first average $\langle ...\rangle$ is performed over all 
SAW configurations on a single backbone of percolation cluster;
the second average $\overline{ \langle ... \rangle}$ is carried out over different realizations of disorder, i.e., over many backbone configurations.

\section{$\Theta$-transition of SASAW on percolative lattice}

\begin{figure}[t!]
 \begin{center}
\includegraphics[width=4.2cm]{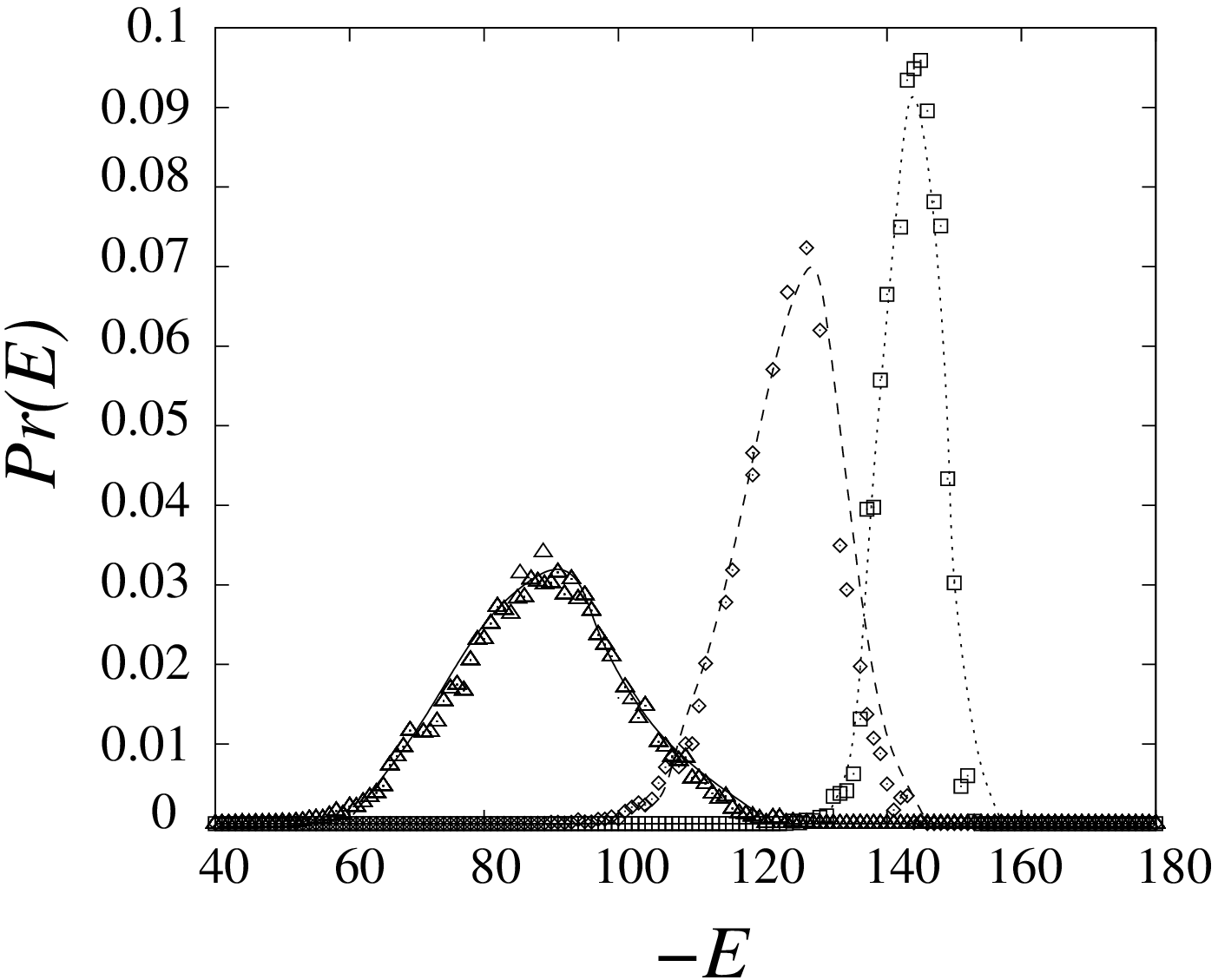}
\includegraphics[width=4.2cm]{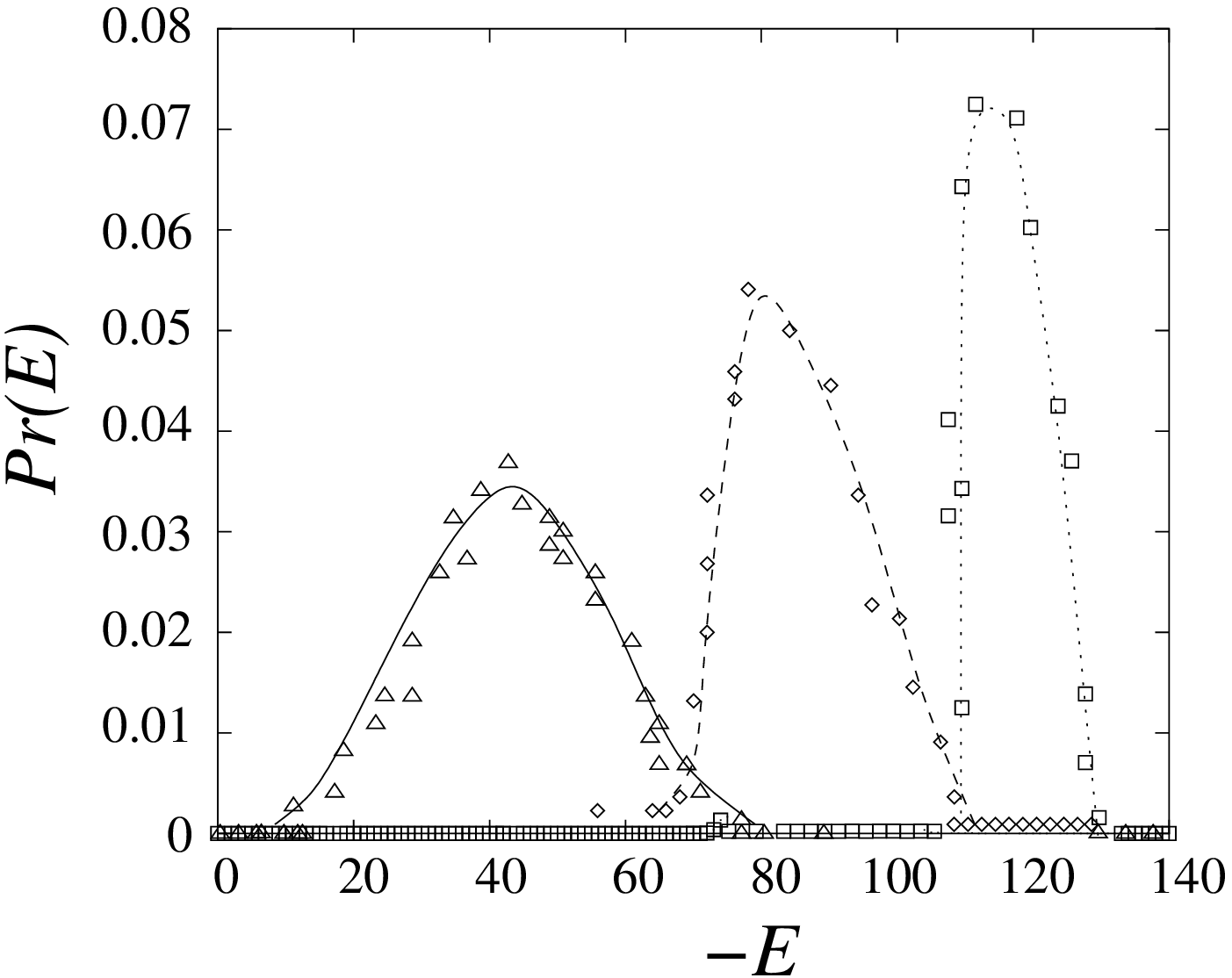}
\end{center}
\caption{\label{enerF0} Left: Energy distribution for $N=180$-step SASAWs on a pure lattice in $d=2$ at different temperatures: $T=0.6$ (squares), $T=1.0$ (diamonds),
$T=1.5$ (triangles). The broadening of the distribution curve emerges close to $T_{\Theta}=1.499\pm0.002$. Right: Energy distribution for $N=180$-step SASAWs 
on the backbone of percolation clusters in $d=2$ at different temperatures: $T=0.2$ (squares), $T=0.4$ (diamonds), $T=0.9$ (triangles).
The broadening of the distribution curve emerges close to $T_{\Theta}^{p_c}=0.92\pm0.02$.}
\end{figure}

Statistical fluctuations of the energy $E$ of a polymer chain, expressed by the behaviour of the specific heat $C_V$, signalize thermodynamic activity in the system,
 and thus the peak structure of $C_V$ as a function of temperature indicates transitions or crossovers between physically different states. In the case
of a polymer system, this corresponds to the transition between globule and coil regimes.
$C_V$ can be expressed via energy fluctuations as follows:
 \begin{equation}
C_V(T)=\frac{1}{T^2}\left(\overline{\langle E^2 \rangle}- \overline{\langle E \rangle^2}\right).
\label{cvperc}
\end{equation} 
It is worthwile first to discuss the behaviour of the energy distribution $Pr(E)$ with varying temperature of the system, as presented in Fig. \ref{enerF0}. 
Since the energy distribution shows one peak only, the transition could be denoted as being second-order like. The width of
the distribution grows with increasing temperature, until it has reached its maximum broadening value. This happens in the vicinity of the $\Theta$-point.   
At higher temperatures, the distribution becomes narrower again.

\begin{figure}[b!]
 \begin{center}
\includegraphics[width=4.2cm]{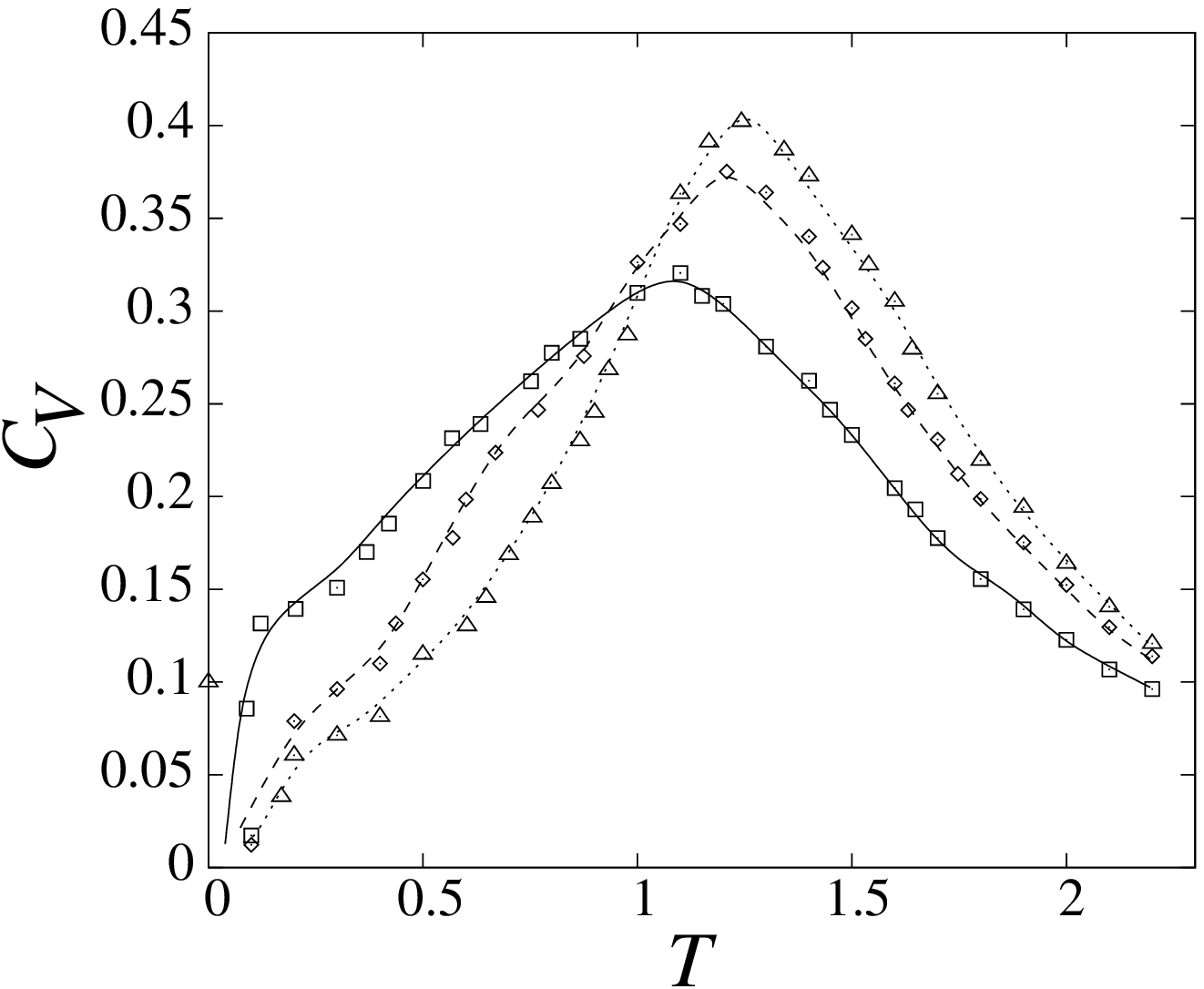}
\includegraphics[width=4.2cm]{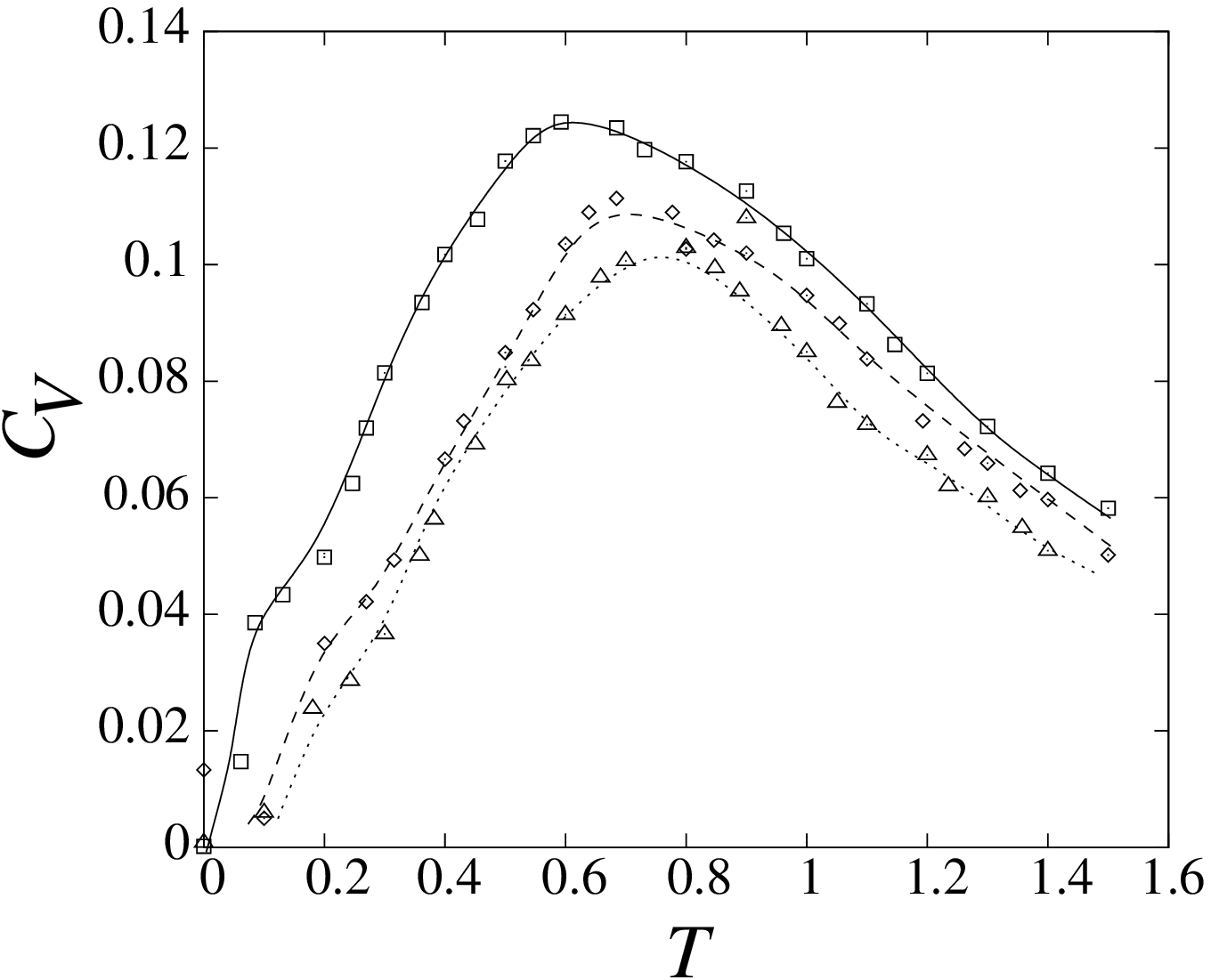}
\end{center}
\caption{\label{cv2F0} Specific heat per monomer as a function of temperature for a SAW in $d=2$ on a pure lattice (left) and backbone of percolation cluster (right).
Squares: $N=50$,  diamonds: $N=100$, triangles: $N=150$.}
\end{figure}
\begin{figure}[t!]
\begin{center}
\includegraphics[width=4.2cm]{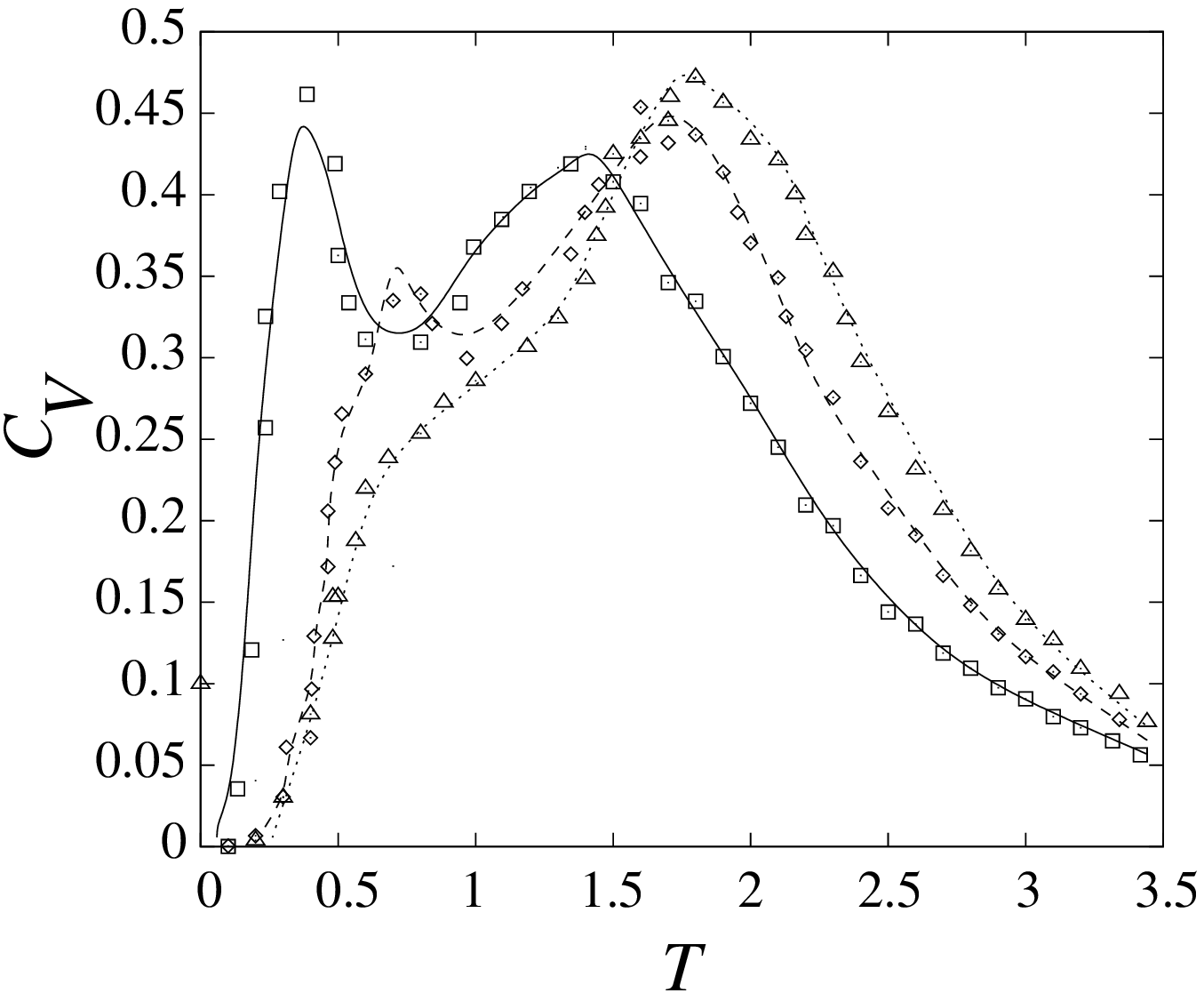}
\includegraphics[width=4.2cm]{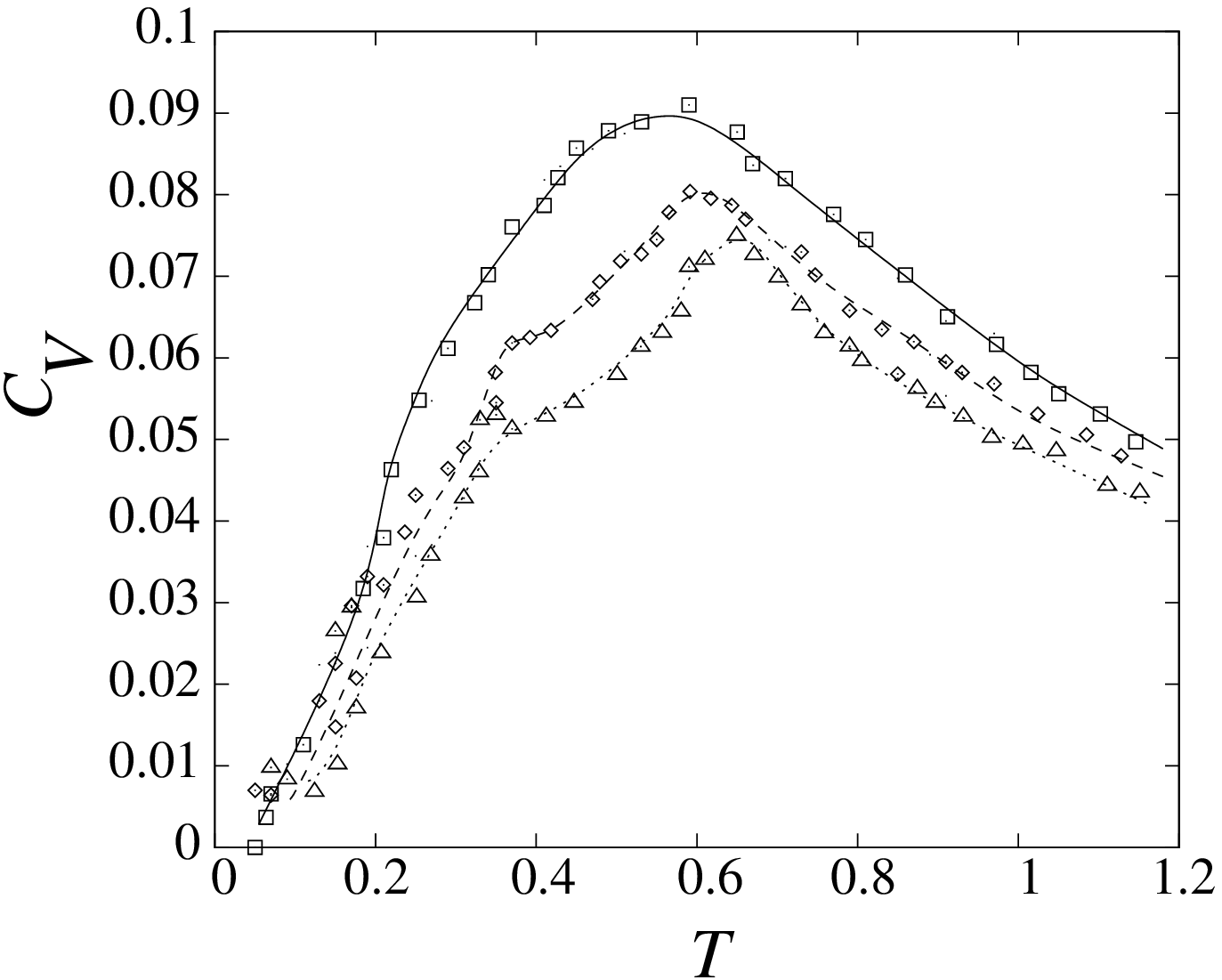}
\end{center}
\caption{ \label{cv3F0} Specific heat per monomer as a function of temperature for a SASAW in $d=3$ on a pure lattice (left) and backbone of percolation cluster (right).
Squares: $N=40$, diamonds: $N=80$, triangles: $N=120$.}
\end{figure}

\begin{figure}[b!]
\begin{center}
\includegraphics[width=4.2cm]{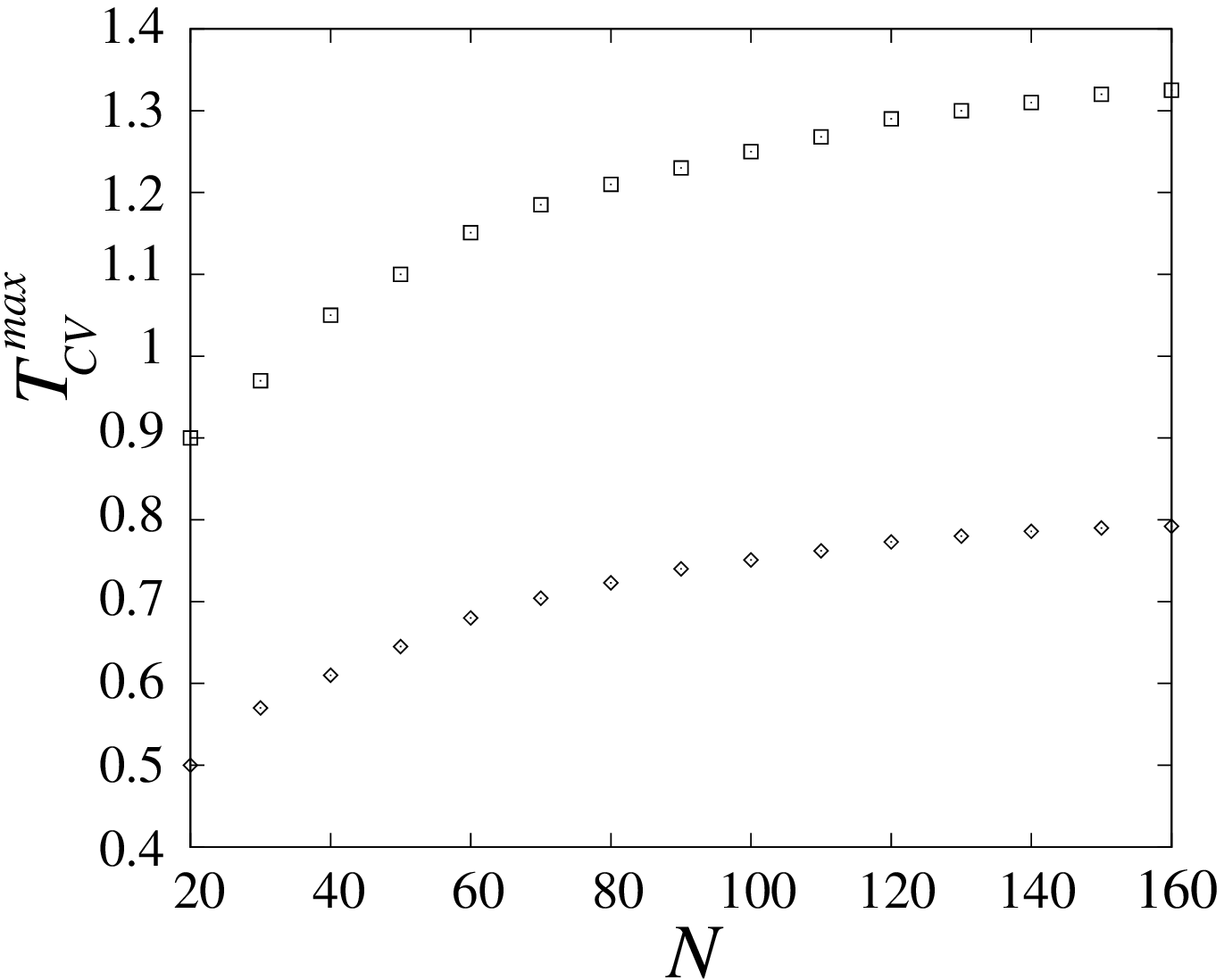}
\includegraphics[width=4.2cm]{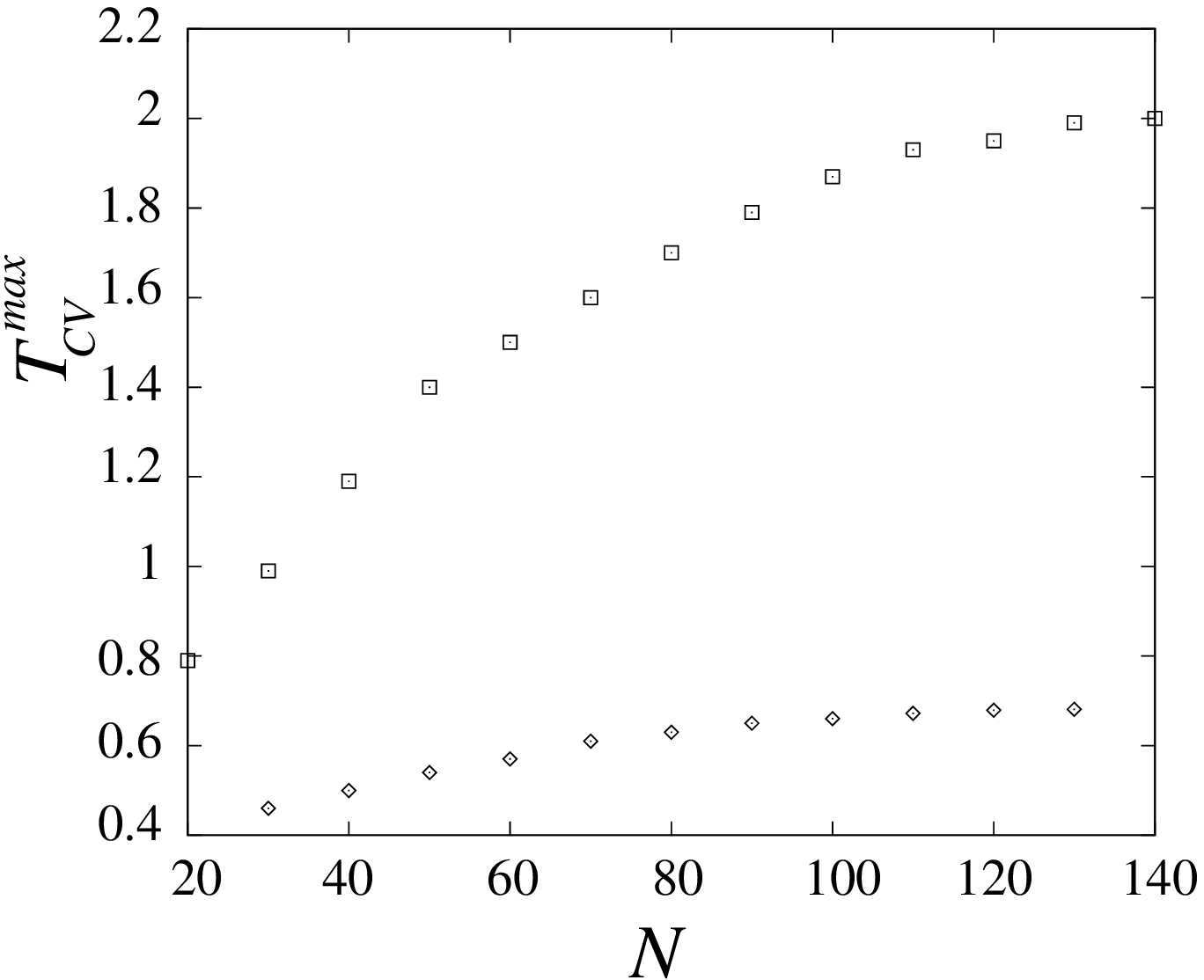}
\end{center}
\caption{ \label{cvmaxF0} Peak temperatures of the specific heat of a SASAW in $d=2$ (left) and $d=3$ (right) at different chain length $N$. Squares: pure lattice, diamondes: backbone of percolation cluster.}
\end{figure}

For comparison with available results and testing our methods, we  performed also simulations on pure lattices. Figures \ref{cv2F0}, \ref{cv3F0} show the  specific-heat behavior of SASAWs on a pure lattice and the backbone of percolation clusters in $d=2,3$ for several different chain lengths. Note, that the maximum of the specific heat per monomer grows with 
$N$ for SAWs on a pure lattice, whereas for the case of the backbone of percolation clusters it descreases with increasing $N$.
 For finite chain length $N$, the temperature defined by position of specific heat maximum $T_{C_V}^{{\rm max}}(N)$ is well below 
 the collapse transition ${\Theta}$-temperature. This finite-size deviation of $T_{C_V}^{{\rm max}}(N)$ from $T_{\Theta}$ obeys scaling behavior with $N$.
Fig. \ref{cvmaxF0} presents the chain-length dependence of the specific-heat peaks for pure and percolative lattices.    
For the case of pure lattice,  $d=3$ is the upper critical dimension for $\Theta$-transition, and the mean-field Flory-Huggins theory \cite{Flory53}  suggests:
\begin{equation}
T_{C_V}^{max}(N)-T_{\Theta}\sim \frac{1}{\sqrt{N}}+\frac{1}{2{N}}.
\end{equation}
In the general case, the approach of $T_{C_V}^{{\rm max}}(N)$ to its limiting value obeys: 
\begin{equation}
{T_{C_V}^{max}(N)}-{T_{\Theta}}\sim a\cdot N^{-\nu_{\Theta}}+\frac{b}{N},
\label{tcv}
\end{equation}
where $a, b$ are constants,  $\nu_{\Theta}$ is the size exponent of a SAW at $\Theta$-point:
\begin{eqnarray}
&&d=2, \mbox{\,pure lattice\!}: \nu_{\Theta}=4/7 \mbox {\,\,\cite{Duplantier82}},\label{phi2}\\ 
&& d=2, \mbox{\,pc\!}:   \nu_{\Theta}^{p_c}=0.74\pm0.02 \mbox {\,\,\cite{Barat93}},\label{phi2pc}\\
&&d=3, \mbox{\,pc\!}: \nu_{\Theta}^{p_c}=0.60\pm0.02 \mbox{\,\,\cite{Barat93}}.\label{phi3pc}
\end{eqnarray}

Note, that we consider the special case of collapse transition on the site-diluted percolative lattice, which was not studied before. 

\begin{figure}[t!]
 \begin{center}
\includegraphics[width=4.2cm]{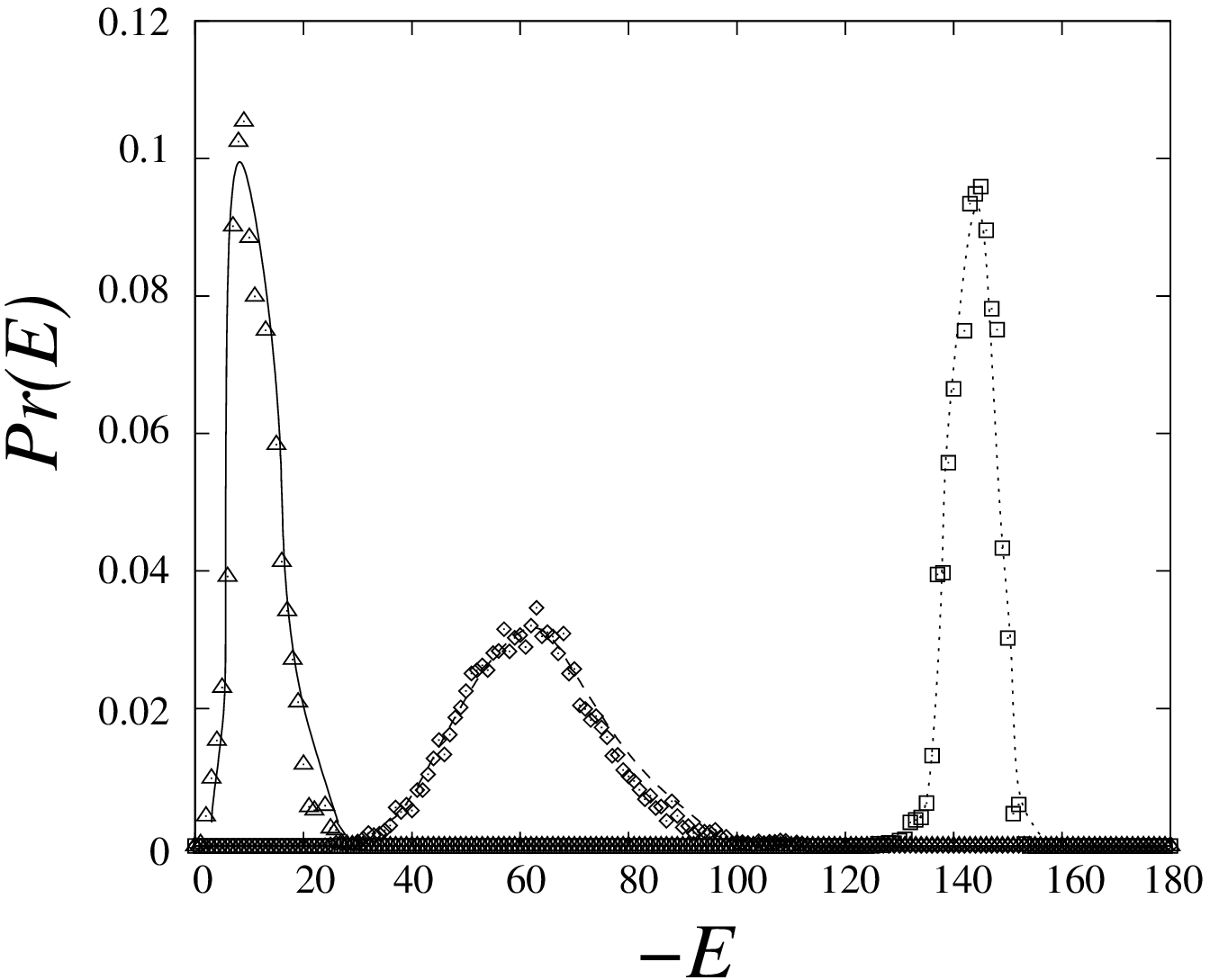}
\includegraphics[width=4.2cm]{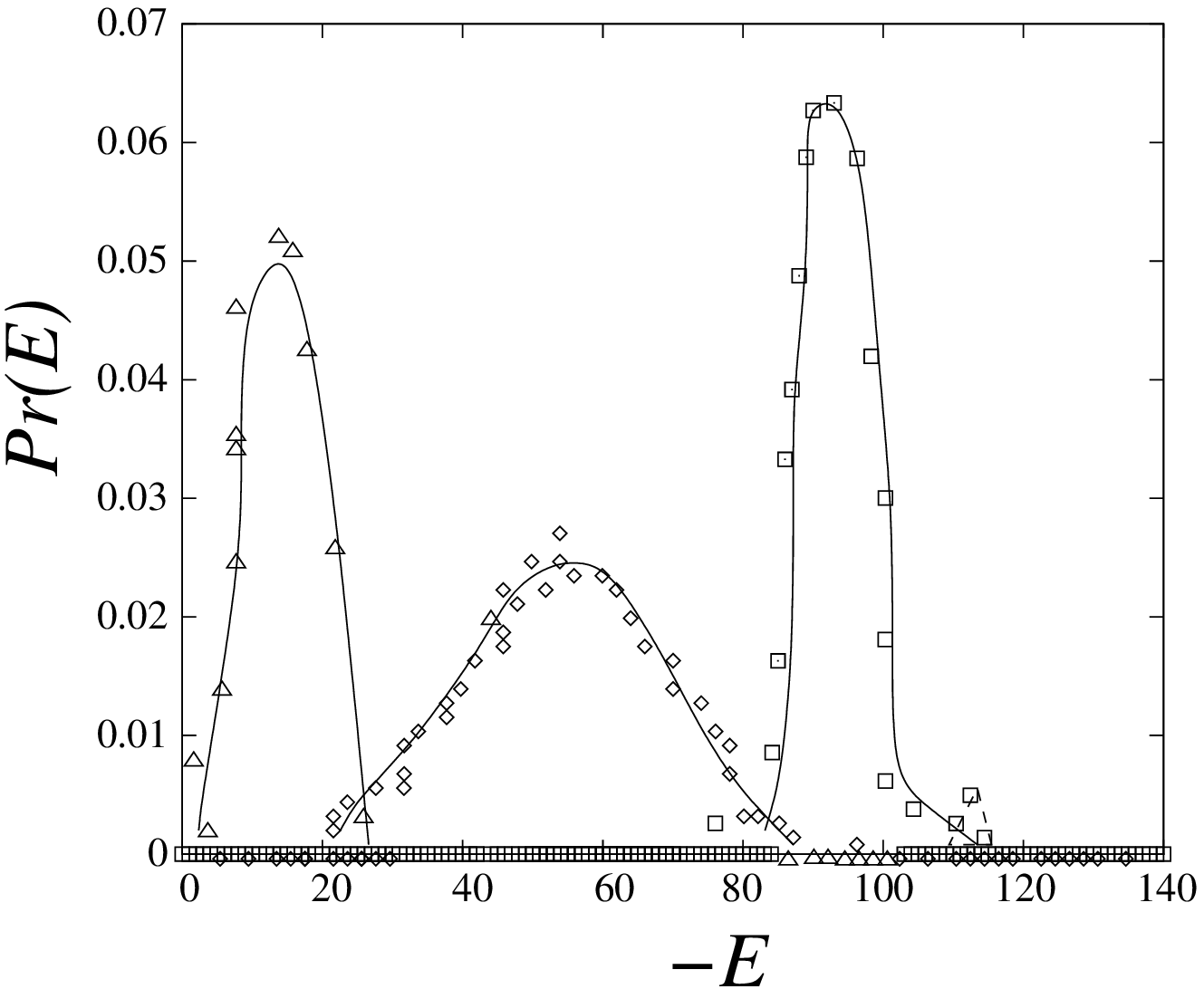}
\end{center}
\caption{\label{enerT} Left: Energy distribution for $N=180$-step SASAWs on pure $d=2$ lattice at $T=0.4$  under a force  $F$ acting in  the environment: $F=0.1$ (squares), $F=1.0$ (diamonds),
$F=1.5$ (triangles). Right: Energy distribution for $N=180$-step SASAWs on the backbone of percolation clusters at $T=0.1$  under force acting in the environment: $F=0.1$ (squares), $F=1.0$ (diamonds),
$F=1.5$ (triangles).}
\end{figure}

 \begin{figure}[b!]
 \begin{center}
\includegraphics[width=4.2cm]{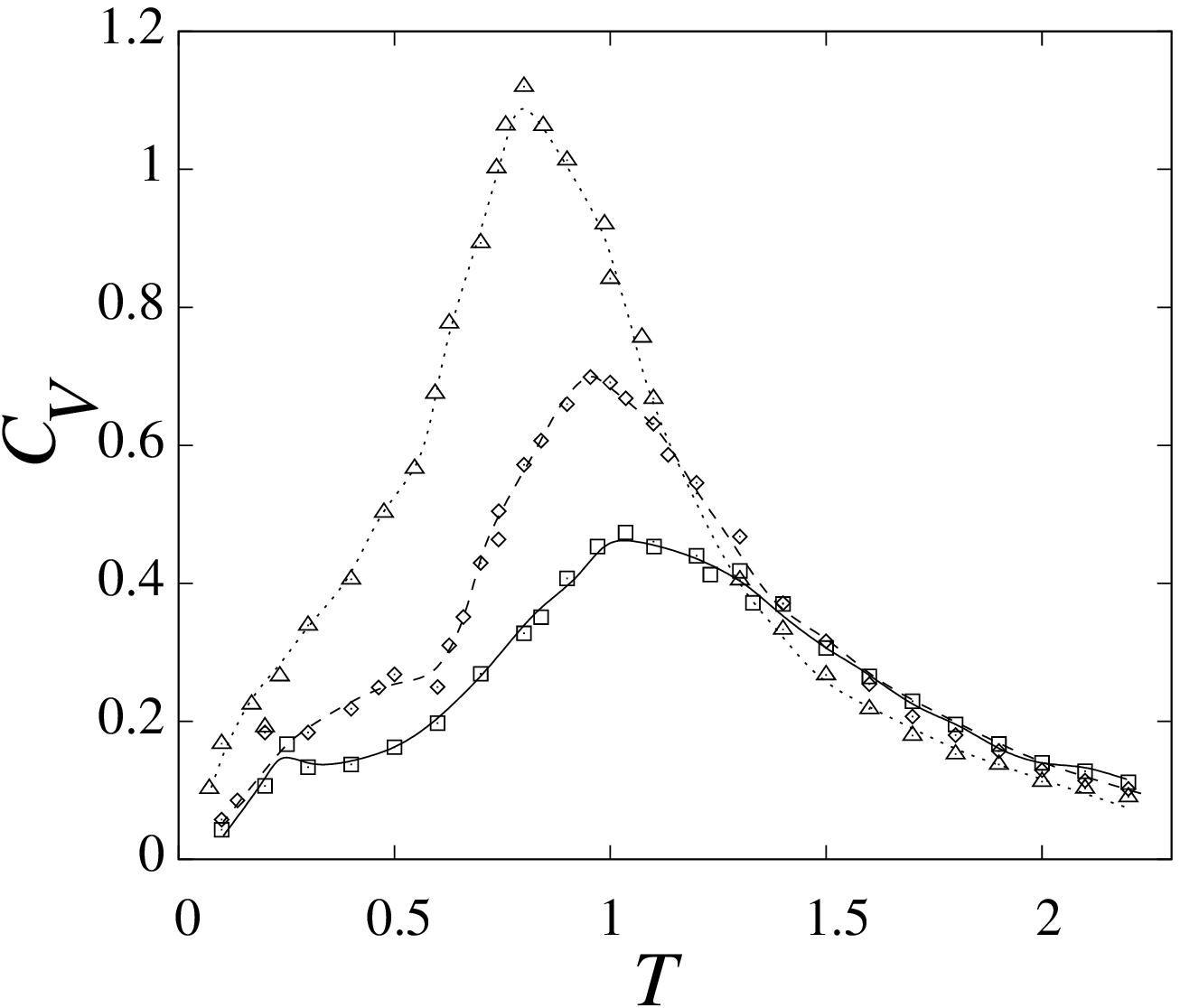}
\includegraphics[width=4.2cm]{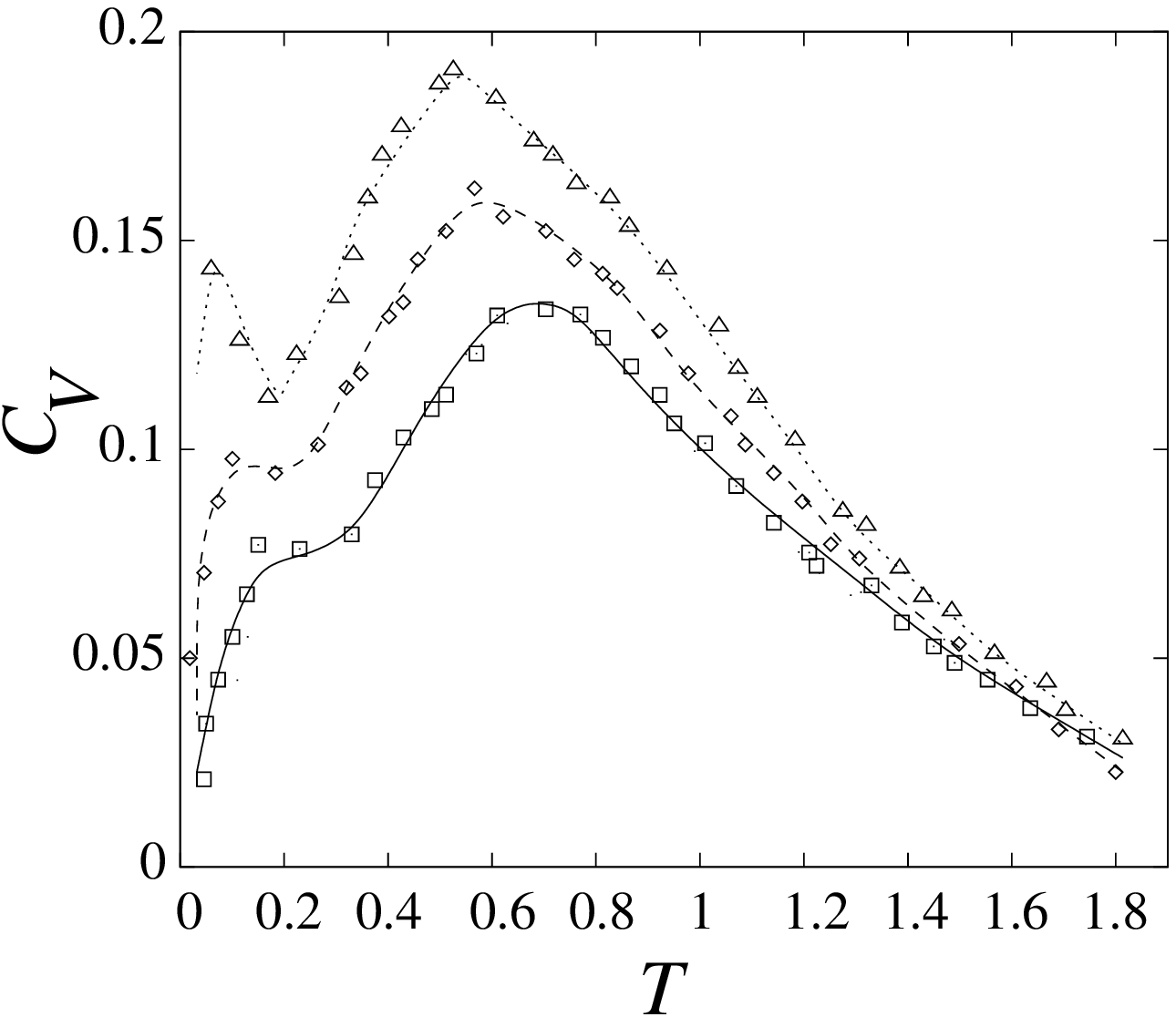}
\end{center}
\caption{\label{cvF2}  Specific heat per monomer of a SASAW with $N=90$ steps in $d=2$ under stretching force $F$ as a function of temperature. Left: pure lattice, right: backbone of percolation cluster. Squares: $F=0.2$, diamonds: $F=0.4$, triangles: $F=0.6$.}
\end{figure}

\begin{figure}[t!]
 \begin{center}
\includegraphics[width=4.2cm]{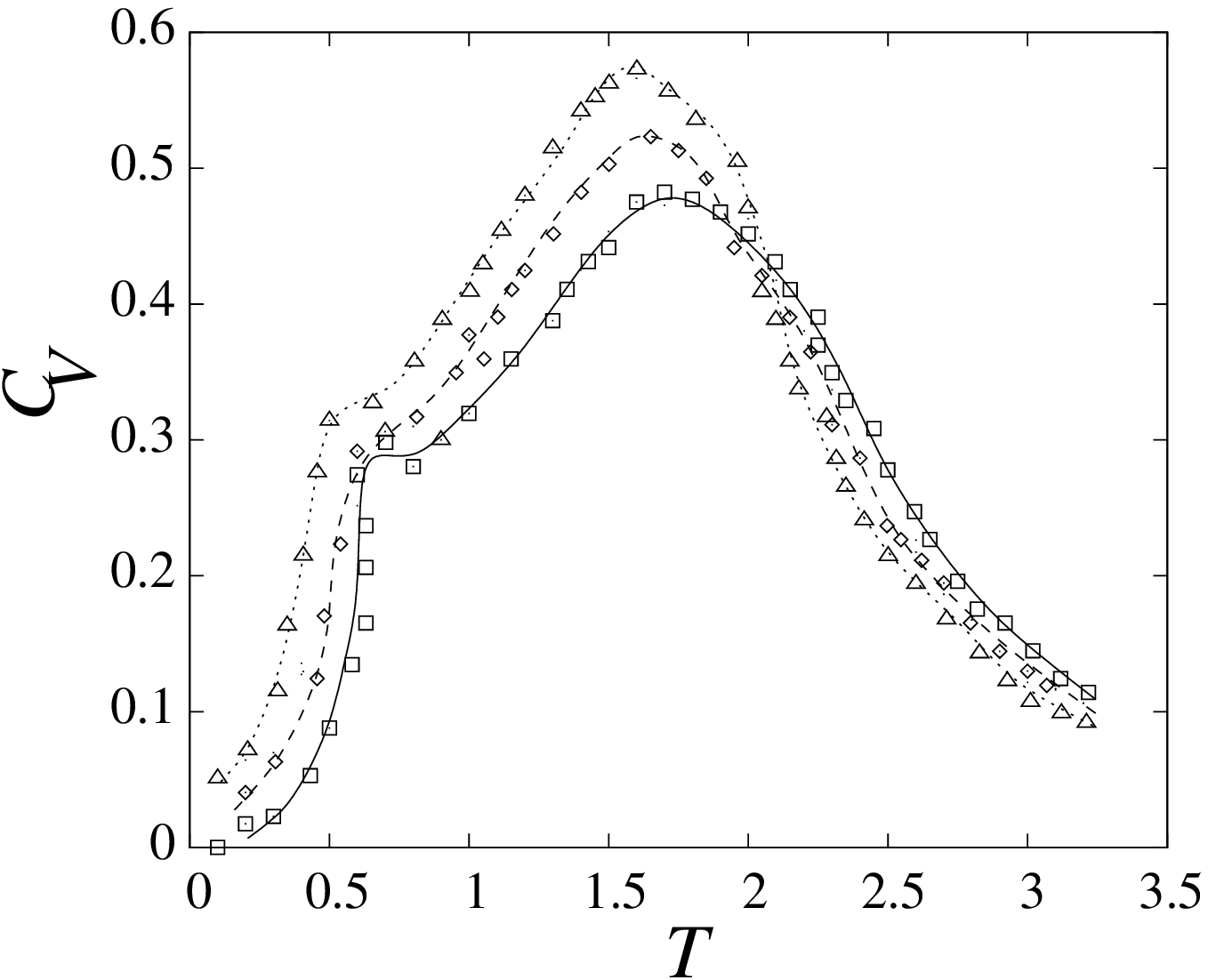}
\includegraphics[width=4.2cm]{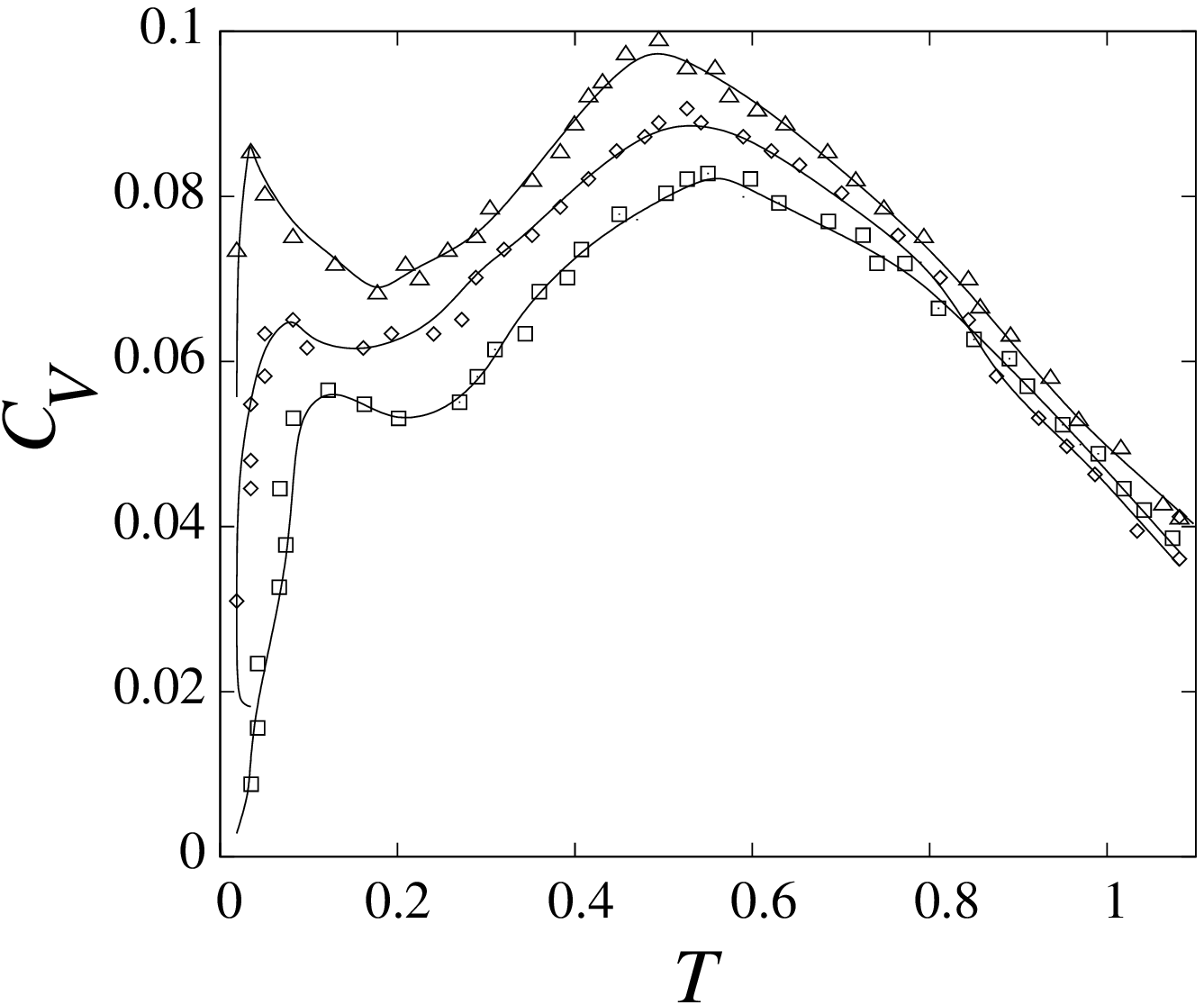}
\end{center}
\caption{\label{cvF3}   
Specific heat per monomer of a SASAW with $N=90$ steps in $d=3$ under stretching force $F$ as a function of temperature. Left: pure lattice, right: backbone of percolation cluster. Squares: $F=0.2$, diamonds: $F=0.4$, triangles: $F=0.6$.}
\end{figure}

Our estimates for $T_{\Theta}^{p_c}$ are obtained by least-square fitting of (\ref{tcv}), taking into account (\ref{phi2pc}) and (\ref{phi3pc}), and presented in Table 1.
As it was already mentioned, the  collapse transition temperature is a non-universal property,  which depends on the lattice type and, in particular, on the connectivity constant. 
As expected, the values of $T_{\Theta}^{p_c}$ appear to be smaller than the pure lattice values, as far as relation $\mu_{p_c}<\mu$ holds. In addition, due 
to the fact that $\mu_{p_c}$ decreases with $d$, the same tendency is reflected in the transition temperature behaviour: 
$T_{\Theta}^{p_c}(d{=}2)>T_{\Theta}^{p_c}(d{=}3)$.
This is in contrast to the values on the pure lattice, but coincides with the tendency of  bond-diluted percolation values, cited in the Introduction. Let us note, finally,
that our values for $T_{\Theta}^{p_c}$ of site percolation  are larger than that of bond-percolation, 
which can be explained again by the difference of connectivity constants of these two fractal lattice structures.   
\section{$\Theta$-transition of SASAW on percolative lattice under stretching force}

Let now act an additional stretching force $F$  in the $x$-direction of SASAW trajectory, fixed at its starting point (see Fig. 1).
The energy distributions $Pr(E)$ at fixed temperature 
and varying  $F$ are presented in Fig. \ref{enerT} for the cases of the pure lattice and the backbone of percolation clusters for comparison. We have chosen the value of $T$  in both cases to be well below the corresponding $\Theta$-temperatures, so that at $F=0$ we restore the energy distribution of the globular state.
 With increasing $F$, the averaged energy of the chain decreases --  applied force stretches the polymer globule. The width of the distribution changes at 
increasing $F$, and reaches its maximum broadening in the vicinity of the $\Theta$-point. The value of the  transition temperature is now shifted by the 
presence of force.  
At higher values of $F$, the distribution becomes narrower again -- the chain is an in extended state.

\begin{figure}[b!]
 \begin{center}
\includegraphics[width=4.3cm]{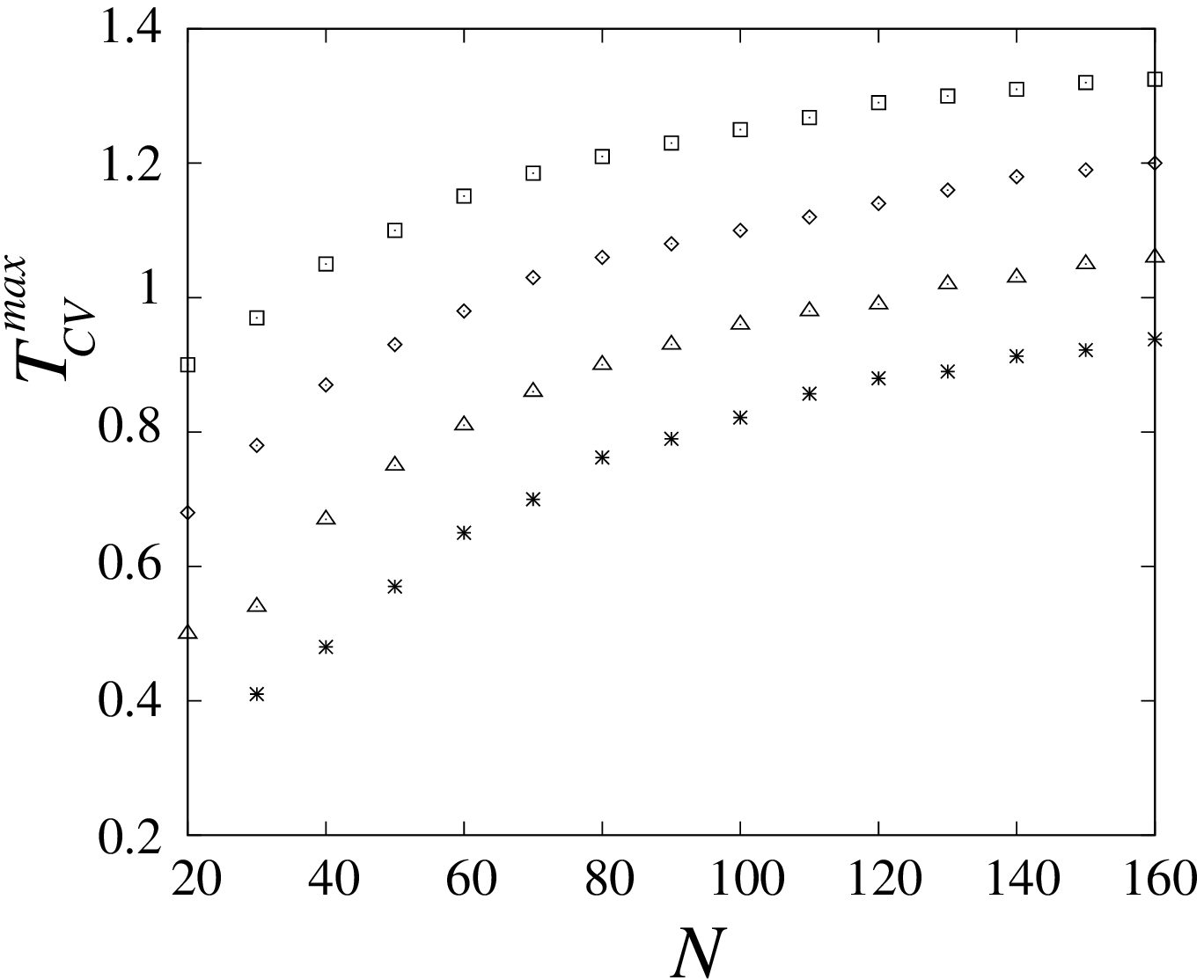}
\includegraphics[width=4.2cm]{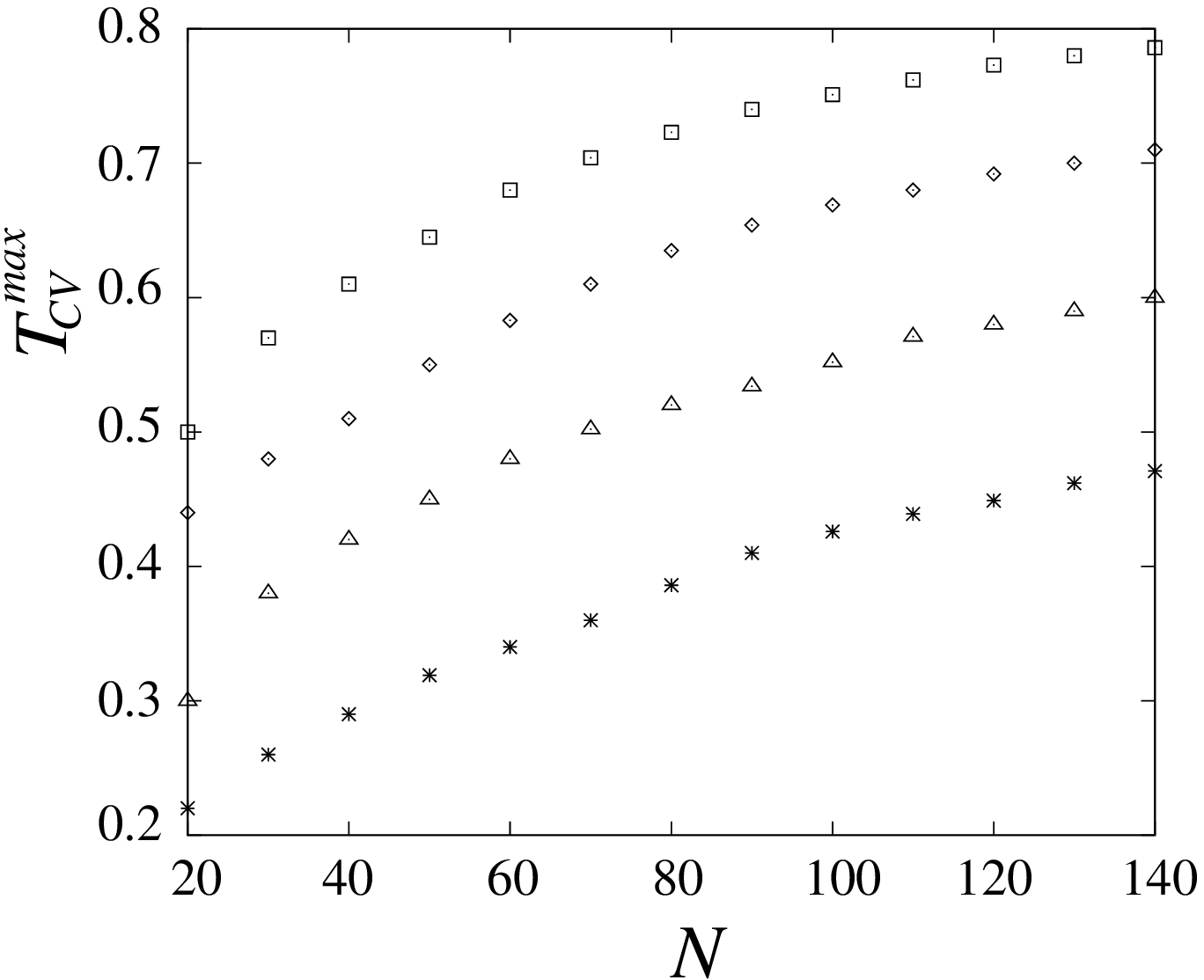}

\includegraphics[width=4.2cm]{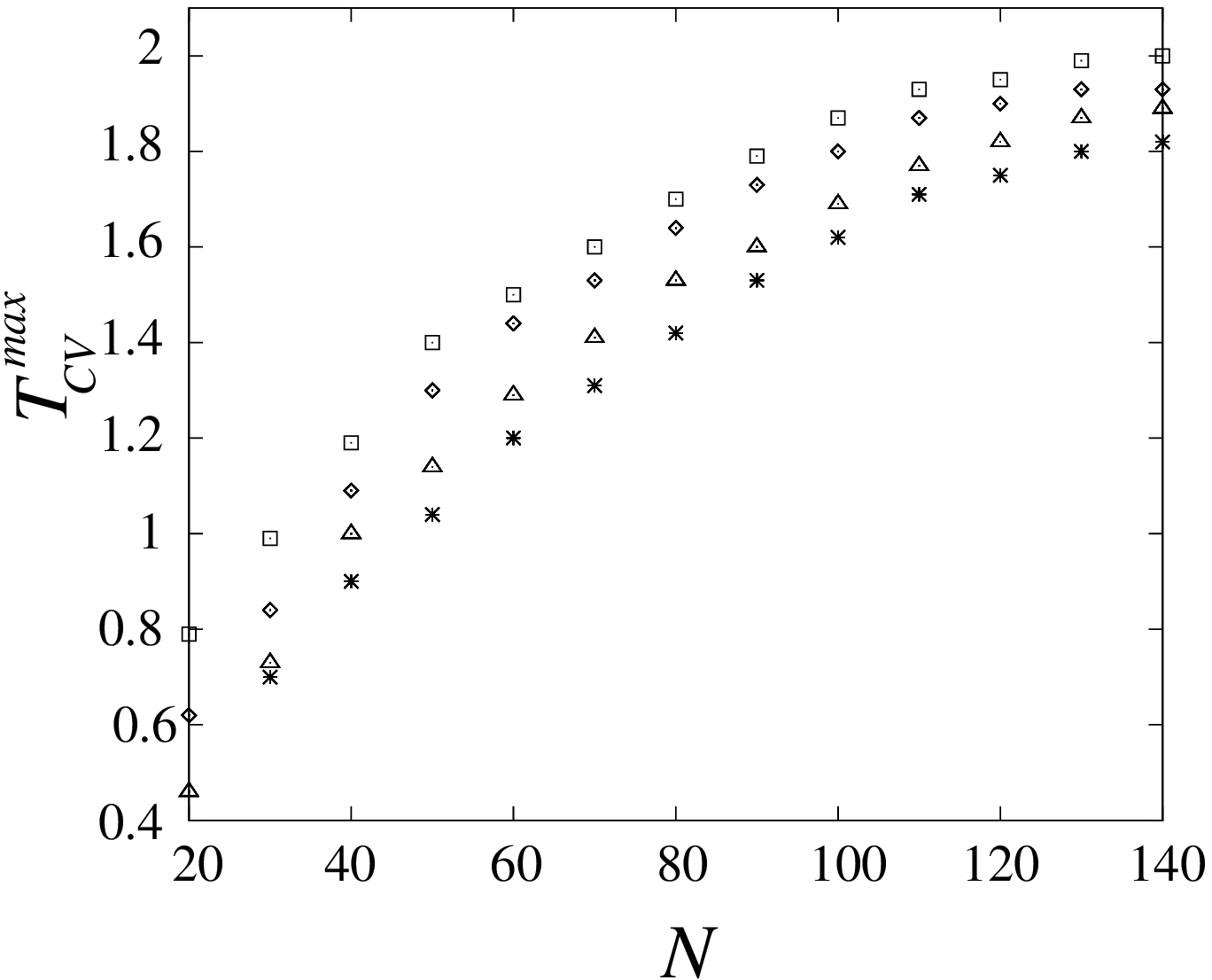}
\includegraphics[width=4.2cm]{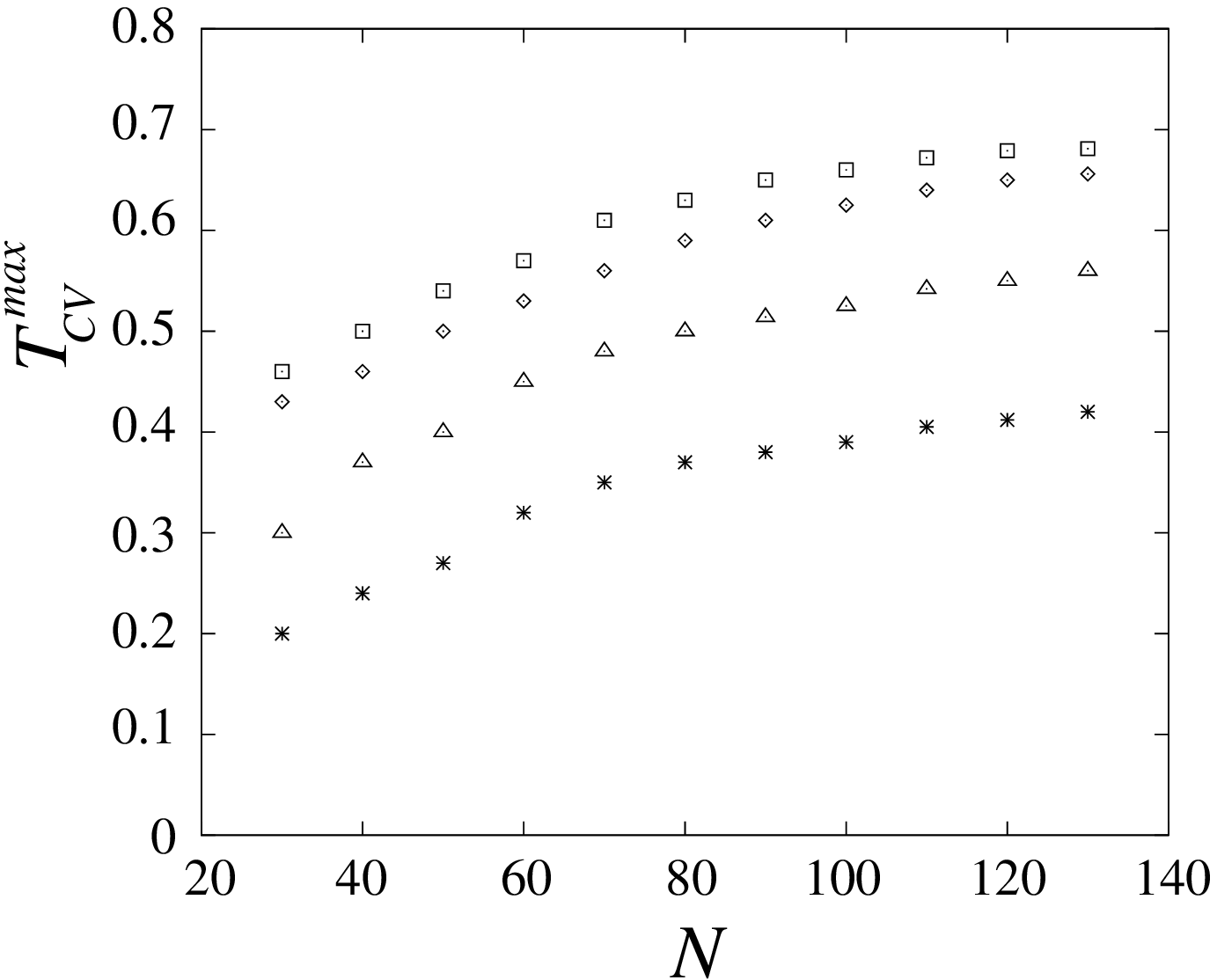}
\end{center}
\caption{\label{cvF} Peak temperatures of the specific heat of a SASAW as a function of chain length $N$. Upper row: $d=2$,  left: pure lattice, right: backbone of percolation cluster.  Lower row: $d=3$, left: pure lattice, right: backbone of percolation cluster.
 Squares: $F=0$, diamonds: $F=0.2$, triangles: $F=0.4$, pluses: $F=0.6$. }
\end{figure}

\begin{figure}[t!]
 \begin{center}
\includegraphics[width=4.2cm]{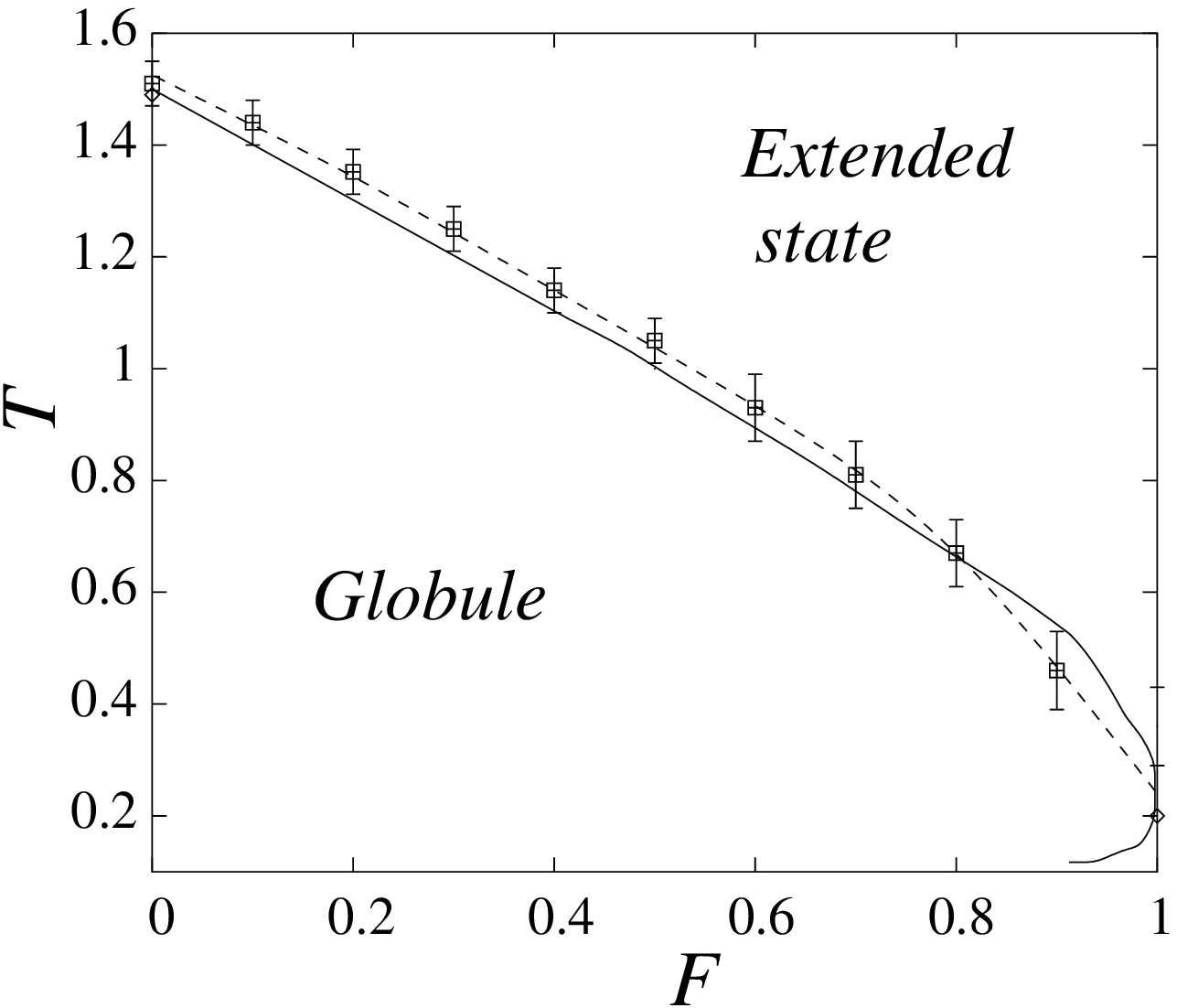}
\includegraphics[width=4.2cm]{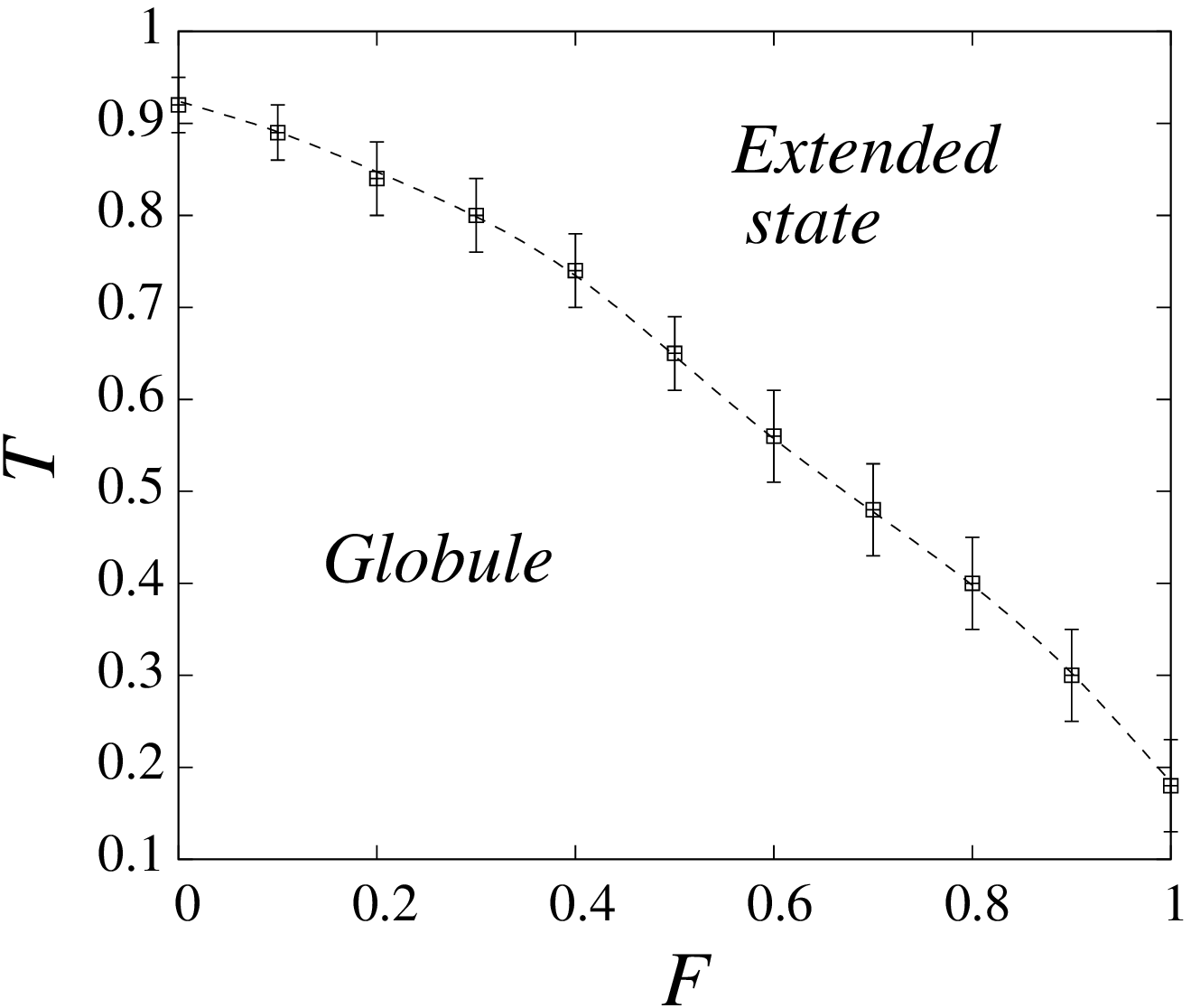}

\includegraphics[width=4.2cm]{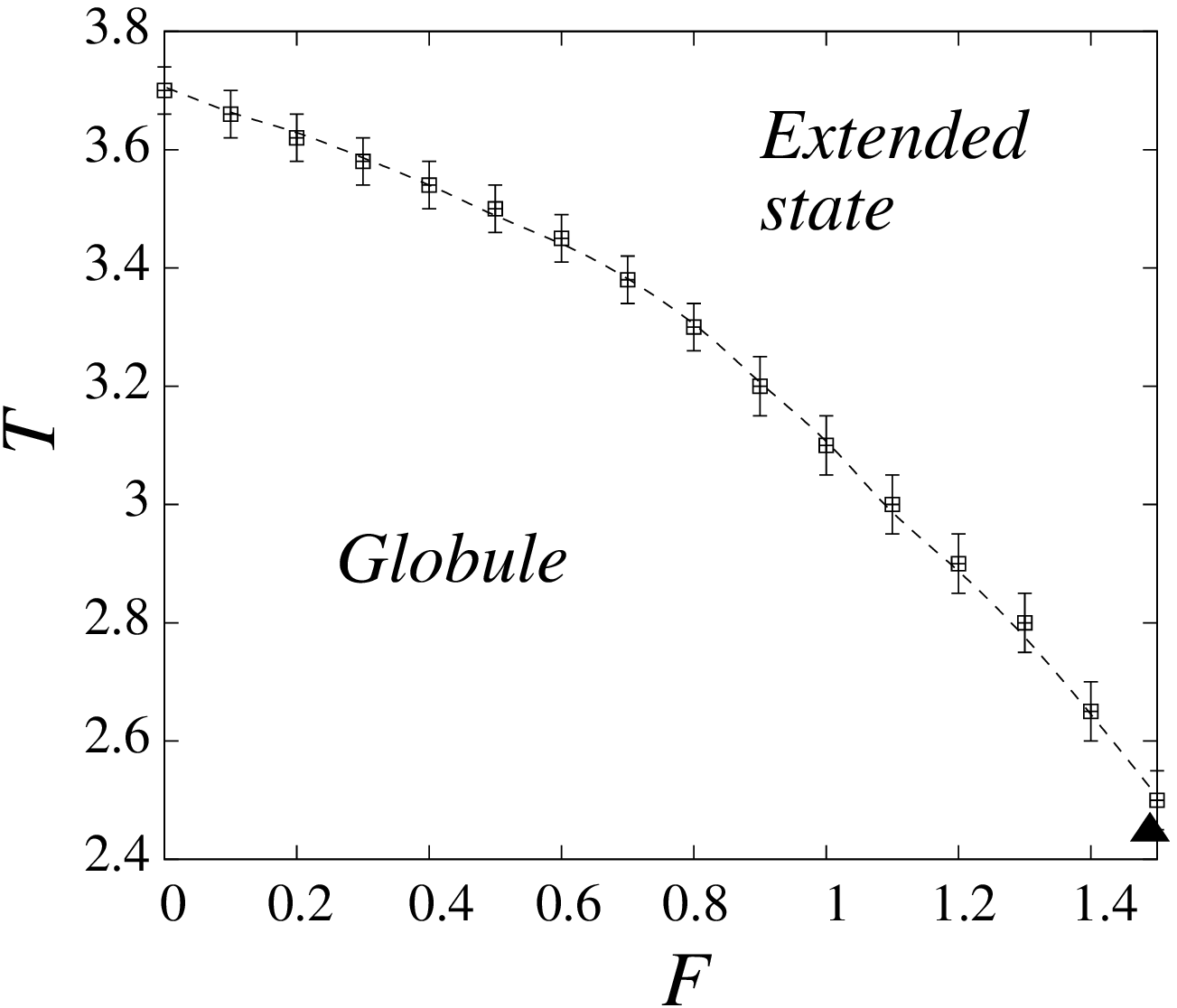}
\includegraphics[width=4.26cm]{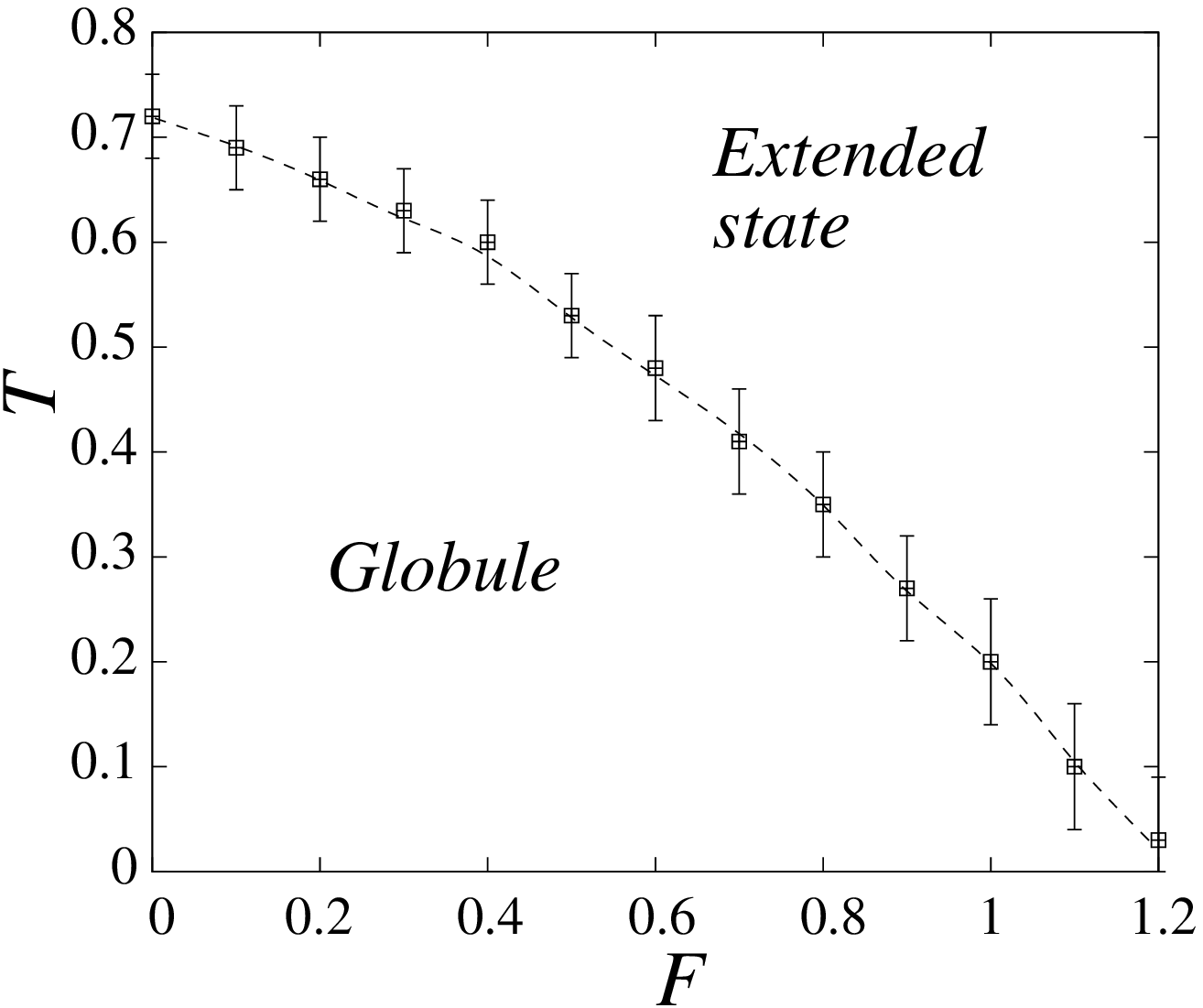}
\end{center}
\caption{\label{cvdiagram} Phase diagrams of the stretching of a SASAW under applied force $F$. Upper row: $d=2$,  left: pure lattice, right: backbone of percolation cluster. The solid line presents results of exact enumeration \cite{Kumar}. Lower row: $d=3$, left: pure lattice, right: backbone of percolation cluster. 
The filled triangle present result of \cite{Grassberger02} $F=1.5, T_{\Theta}\approx 2.46$. }
 \end{figure}
\begin{table}[b!]
\small{
\caption{Values of $\Theta$-temperature on the backbone of percolation clusters in $d=2, 3$ at varying force $F$.}
\label{tab2}
\begin{center}
\begin{tabular}{lcccccr}
\hline  
\hline
$ F$ & 0 & 0.2 & 0.4 & 0.6 & 0.8 & 1.0 \\ 
\hline 
$T_{\Theta}^{p_c}(d=2)$  &  0.92(2) & 0.84(2) & 0.74(3) & 0.56(2) & 0.40(4) & 0.18(5)\\ 
$T_{\Theta}^{p_c}(d=3)$   & 0.71(2)& 0.66(2)& 0.60(2)& 0.48(3) &  0.35(2)& 0.20(2) \\
 \hline 
\hline
\end{tabular}
\end{center}
}
\end{table}
 To study the $\Theta$-transition of SASAWs, when the external stretching force is acting in the environment, we are working in the ``constant-force" ensemble.
Fixing the value of $F$,  we study the specific-heat behavior (Figs. \ref{cvF2}, \ref{cvF3}). Analyzing the peak structure of the specific heat, we immediately conclude,
that  increasing the value of $F$ leads to decreasing the transition temperature. Fig. \ref{cvF} presents the chain-length dependence of the specific-heat peaks for a pure and percolative lattices at several different values of $F$.
  
For finite chain length $N$, the temperature defined by the position of the specific-heat maximum $T_{C_V}^{{\rm max}}(N)$ is well below 
 the collapse transition ${\Theta}$-temperature. 
Our estimates for $T_{\Theta}^{p_c}$ in the presence of force are obtained by least-square fitting of (\ref{tcv}) with (\ref{phi2}), (\ref{phi2pc}), (\ref{phi3pc}). 
 Results are presented in Fig. \ref{cvdiagram} in the form of a phase diagram of transitions from globule to extended state, and  listed in Table 2.
For the pure lattice in $d=2$, we compare our results with an exact enumeration study of Kumar et al. \cite{Kumar}, where SASAWs of length up to $N=55$ under stretching force were studied. Our results also appear to be in good correspondence with that of Ref. \cite{Marenduzzo02}.

\begin{figure}[t!]
 \begin{center}
\includegraphics[width=4cm]{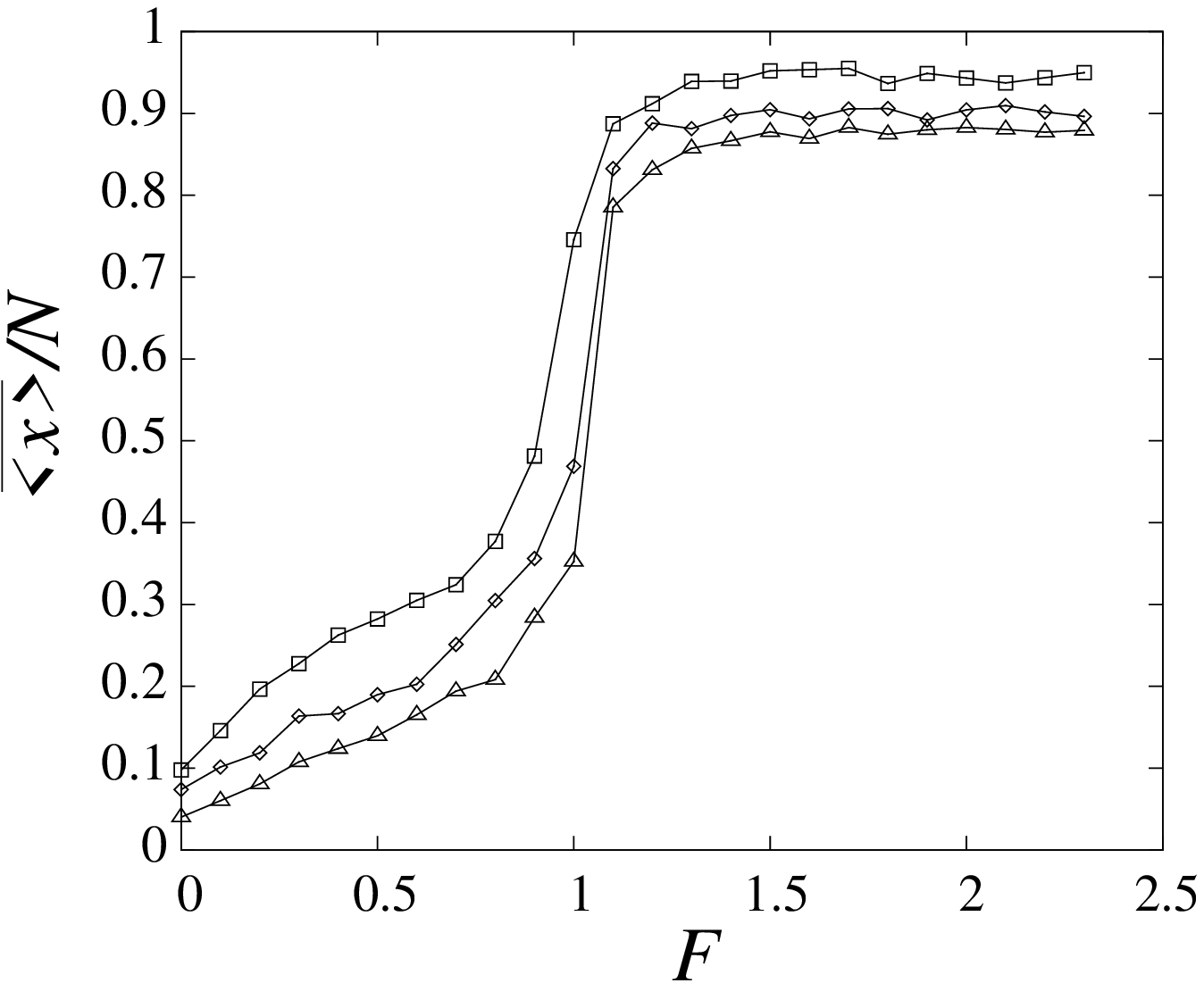}
\includegraphics[width=4cm]{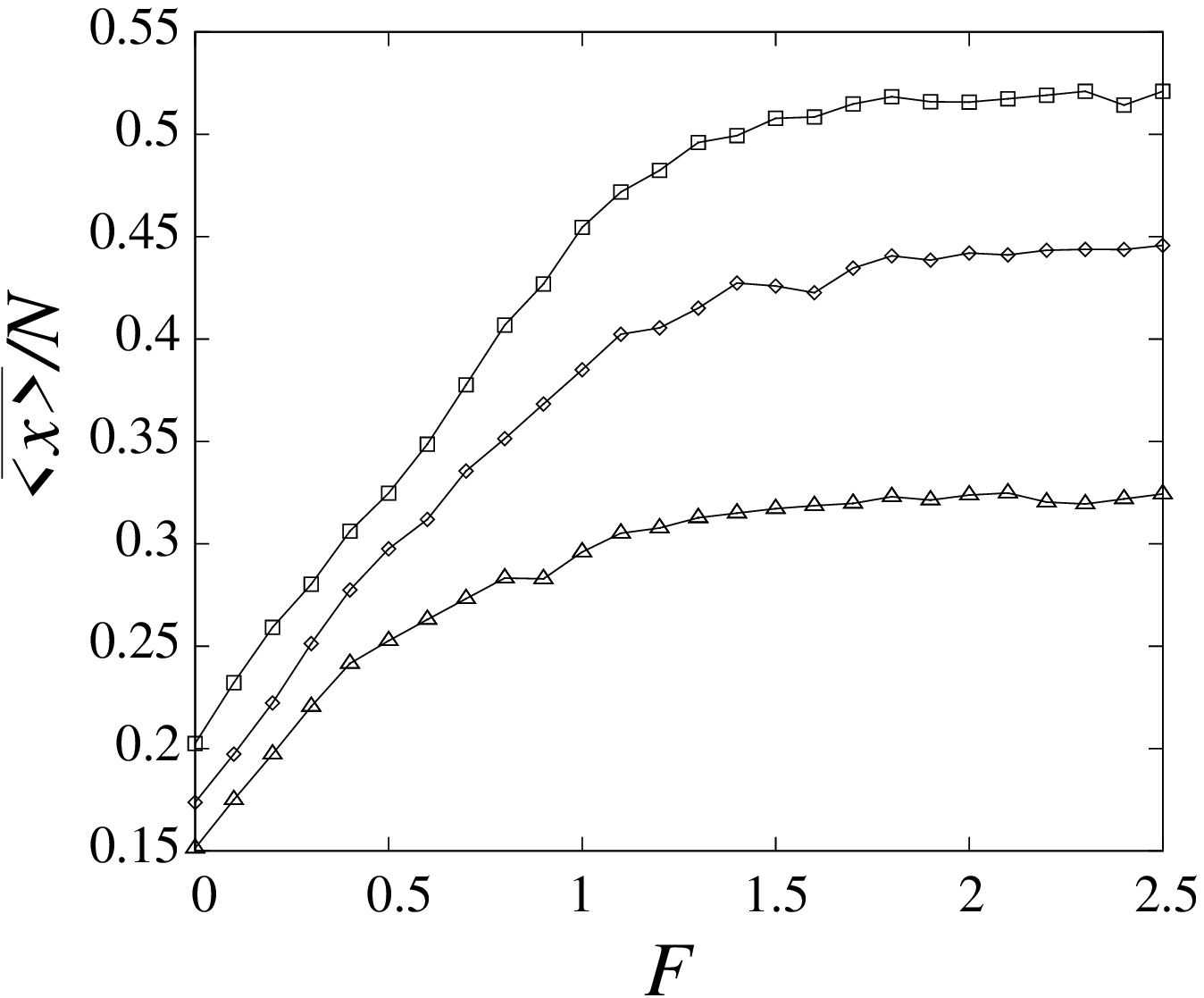}
\includegraphics[width=4cm]{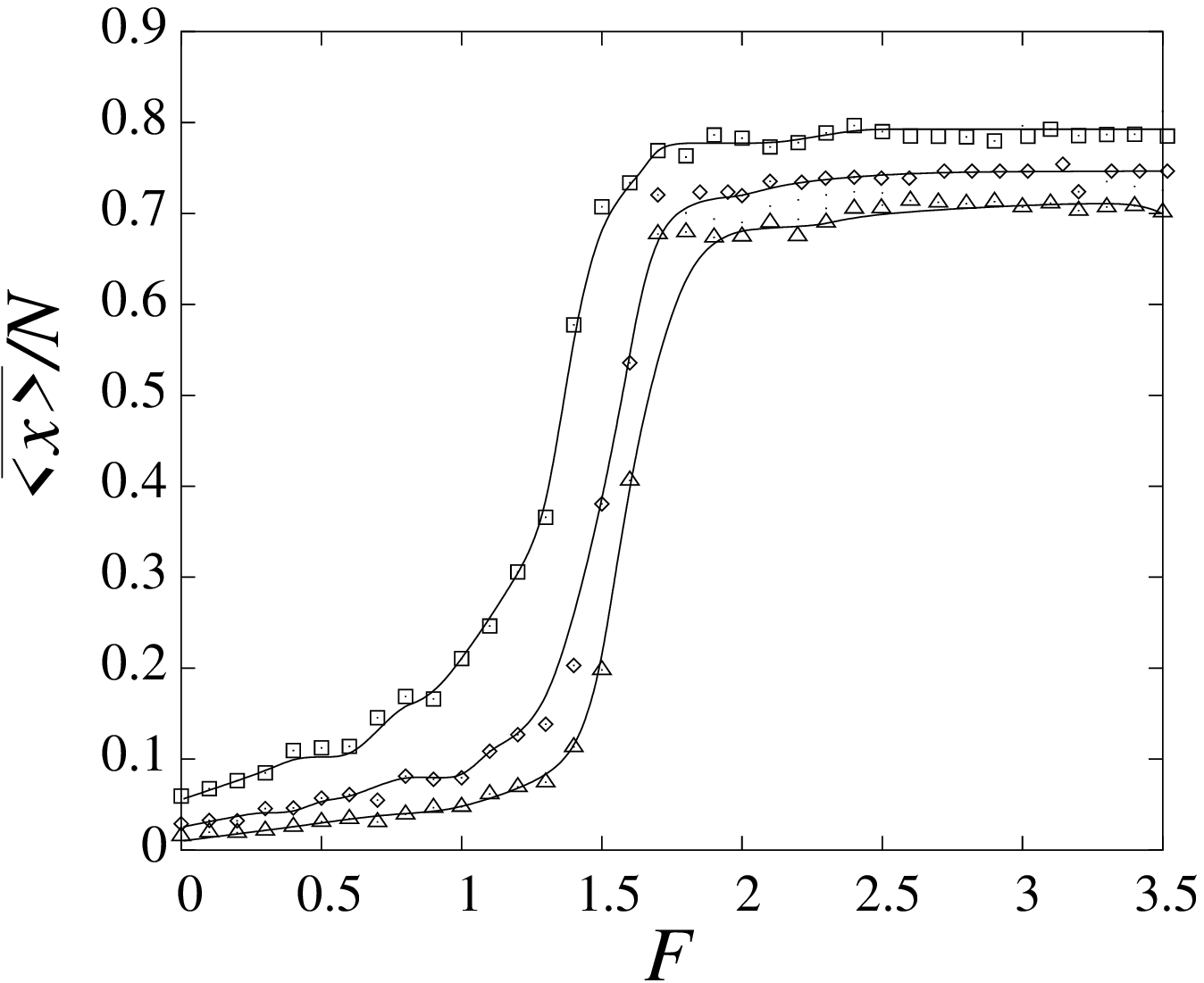}
\includegraphics[width=4cm]{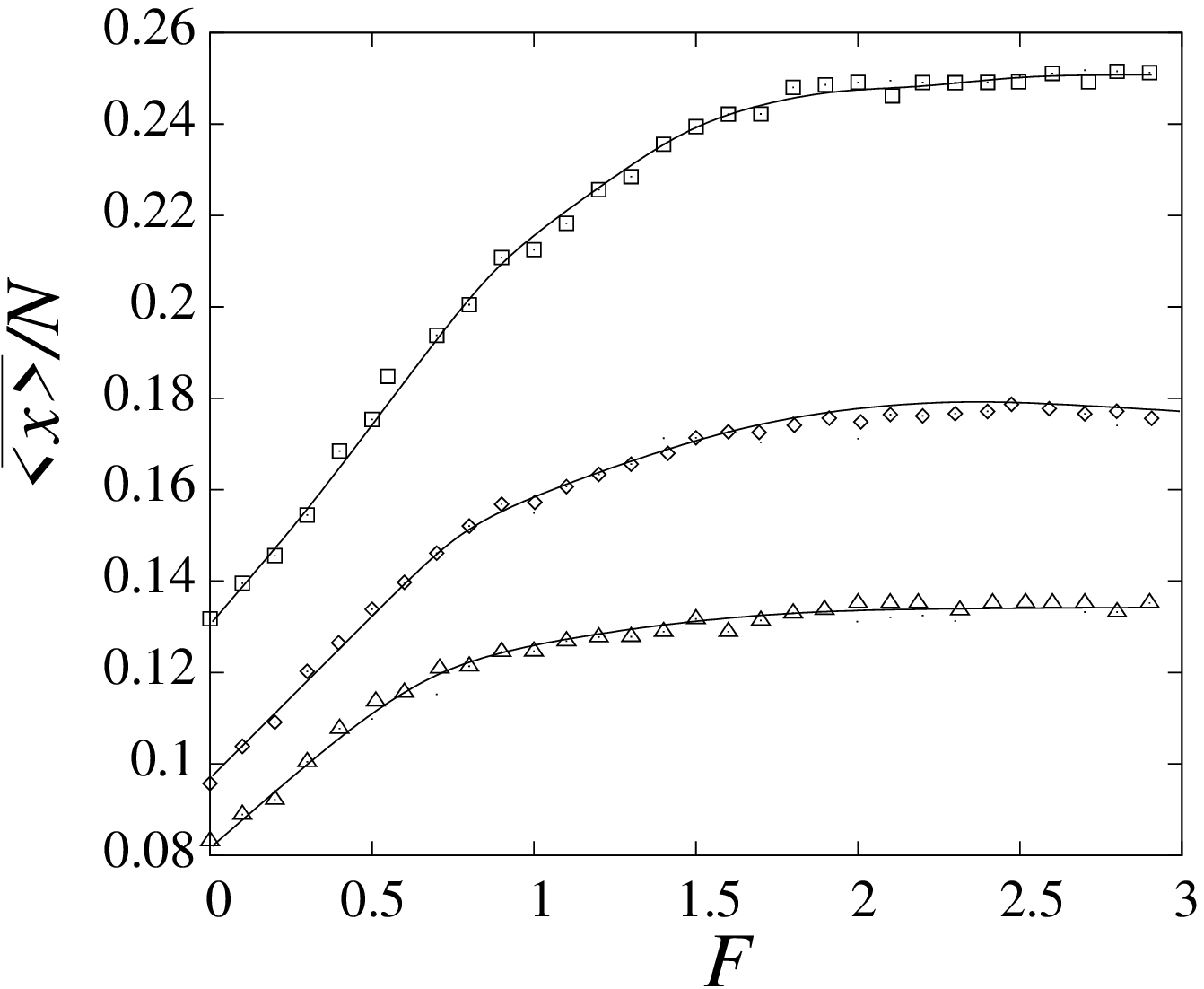}
\end{center}
\caption{\label{exten} Upper row: Averaged extentions of a SASAW under applying force acting in a $d=2$ environment at $T=0.2$; left - pure lattice, right - backbone of percolation cluster, 
squares: $N=20$, diamonds: $N=50$, triangles: $N=100$.
Lower row: Averaged extentions of a SASAW under applying force acting in a $d=3$ environment; left - pure lattice ($T=1.8$), right - backbone of percolation cluster ($T=0.4$), 
squares: $N=20$, diamonds: $N=50$, triangles: $N=100$.
}
\end{figure}

Averaged extention in $x$-direction, giving information about the internal structure of the polymer configuration under applying force, is presented in Fig. \ref{exten} for the cases of pure lattice and backbone of percolation cluster. At small forces, a polymer chain is still in the compact folded state and slightly oriented along the force direction. At larger forces, the polymer chain has a conformation similar to the extended (swollen) structure. Note, that completely stretched states, corresponding to ${\overline {\langle x \rangle}} /N \simeq 1$ can be obtained only on the pure case and are not accessible on the percolative lattices due to complicated fractal structure on the underlying percolation cluster. Our estimates of $ {\overline {\langle x \rangle}/N}$ in $d=2$ appear to be in a good correspondence with that of available exact enumeration studies \cite{Kumar,Kumarpc}.

\section{Conclusions}
We studied the self-attracting self-avoding walks on disordered lattices 
in space dimensions $d=2$, $3$, modelling flexible polymer macromolecules in porous environment. 
We considered the special case, when the concentration of disorder is 
exactly at the percolation threshold, so that an incipient percolation cluster of sites, allowed for SAWs, emerges on the lattice. 
In our study, SASAWs reside only on the backbone of the percolation cluster, which has a fractal structure. 

In the first part of our study, attention has been paid to the influence
 of the fractal structure of the underlying lattice on properties of the coil-globule transition. Applying the pruned-enriched Rosenbluth algorithm, 
we obtain estimates
of the collapse transition temperature $T_{\Theta}^{p_c}$ of SASAWs on  site-diluted percolative lattices 
in $d=2$ and $3$ dimensions (note, that so far only estimates for the 
bond-percolation case have been found).
The values of $T_{\Theta}^{p_c}$ appear to be smaller than the pure lattice value. This can be explained, remembering that the collapse 
transition temperature is proportional to the connectivity constant $\mu$ of SAWs on a given lattice. In addition, due 
to the fact that $\mu_{p_c}$ decreases with $d$, the same tendency is reflected in the anomalous transition temperature behaviour: 
$T_{\Theta}^{p_c}(d{=}2)>T_{\Theta}^{p_c}(d{=}3)$.

Next, keeping one end of a SASAW trajectory on the backbone of percolation cluster fixed,  
we applied a stretching force, acting in some chosen direction (say, $x$).
Especially interesting was to study this problem in $d=3$, which corresponds to real polymer systems and was not considered before.
We estimated the shift of the $\Theta$-temperature of the globule-coil transition 
under the stretching and constructed phase diagrams of collapsed and extended states coexistence in $d=2,3$. As expected, the presence 
of stretching force in environment leads to a decreasing $\Theta$-temperature value. 

\section*{Acknowledgement}
 We thank Sanjay Kumar for interesting discussions and communicating Ref.~\cite{Kumarpc} prior to publication. 
V.B. is grateful for support through the Alexander von Humboldt Foundation and the S\"achsische DFG-Forschergruppe FOR877.

\end{document}